\documentclass[letterpaper,english,aps,pre,reprint]{revtex4-1}
\pdfoutput=1
\usepackage[T1]{fontenc}
\usepackage[latin9]{inputenc}
\usepackage{babel}
\usepackage{amstext,amsmath}
\usepackage{graphicx}
\usepackage{bm}
\usepackage{xr}
\usepackage[breaklinks=true,pdfborder={0 0 1},colorlinks=true,citecolor=blue,filecolor=blue,urlcolor=blue]{hyperref}
\usepackage[usenames]{color}

\makeatletter

\pdfpageheight\paperheight
\pdfpagewidth\paperwidth

\providecommand{\tabularnewline}{\\}
\newcommand{\lyxdot}{.}

\newcommand{\kT}{k_\text{B}T}
\newcommand{\fa}{f_\text{a}}
\newcommand{\lcg}{b}
\newcommand{\ra}{\rho_\text{a}}
\newcommand{\rd}{\rho_\text{p}}
\newcommand{\Daa}{D_\text{aa}}
\newcommand{\Ddd}{D_\text{pp}}
\newcommand{\Dad}{D_\text{ap}}
\newcommand{\Dda}{D_\text{pa}}
\newcommand{\Fda}{\mathbf{F}_\text{ap}}
\newcommand{\Jad}{\mathbf{J}_\text{ap}}

\definecolor{Blue}{rgb}{0,0.0,1.0}
\definecolor{Red}{rgb}{1.0,0.0,0.0}
\definecolor{Green}{rgb}{0.0,0.35,0.0}
\definecolor{Grey}{rgb}{0.5,0.5,0.5}

\makeatother

\begin{document}

\title{Spontaneous Segregation of Self-Propelled Particles with Different Motilities}

\author{Samuel R. McCandlish}

\author{Aparna Baskaran}

\author{Michael F. Hagan}
\email{hagan@brandeis.edu}

\affiliation{Martin A. Fisher School of Physics, Brandeis University, Waltham,
MA, 02454}

\date{\today}
\begin{abstract}
We study mixtures of self-propelled and passive rod-like particles in two
dimensions using Brownian dynamics simulations. The simulations demonstrate
that the two species spontaneously segregate to generate a rich array of
dynamical domain structures whose properties depend on the propulsion
velocity, density, and composition. In addition to presenting phase diagrams
as a function of the system parameters, we investigate the mechanisms
driving segregation. We show that the difference in collision frequencies
between self-propelled and passive rods provides a driving force for
segregation, which is amplified by the tendency of the self-propelled rods to
swarm or cluster. Finally, both self-propelled and passive rods exhibit
giant number fluctuations for sufficient propulsion velocities.

\end{abstract}
\maketitle

\section{Introduction}

Systems of self-propelled particles are found in nature on many scales,
ranging from flocks of birds to colonies of bacteria
and assemblies of cytoskeletal filaments \cite{Julicher2007a,Koenderink2009a,Liu2008b,Mizuno2007a,Schaller2010a,Swaminathan2008a,Toner1998,Zhang2010a}. Theoretical and experimental investigations of propelled or otherwise active particles have established a remarkable array of emergent behaviors not possible in passive systems, such as giant number fluctuations \cite{Ramaswamy2003,Zhang2010a,Chate2008,Deseigne2010a,Toner2005,Narayan2007a,Chate2008,Schaller2010a,Chate2006,Kudrolli2008a}, spontaneous flow \cite{Giomi2011,Giomi2008}, and unconventional rheological properties \cite{Giomi2008}.  Of particular relevance to our work, self-propelled particles have been shown to exhibit dramatic swarming or clustering  \cite{Baskaran2008c,Bertin2006,Bertin2009,Chate2008,Chate2008,Cisneros2011a,Deseigne2010a,Ginelli2010a,Head2010a,Jia2008a,Kudrolli2008a,Narayan2007a,Peruani2006a,Peruani2011,Peruani2011a,Saintillan2011,Schaller2010a,Toner1998,Vicsek1995a,Yang2010a,Zhang2010a}. These unconventional behaviors arise because active particles consume or dissipate energy to fuel internal changes that generate motion. Inevitably, some members of an active system lose this ability by breaking or dying, or fail to acquire it to begin with. In colonies of motile bacteria, for example, there is typically a significant fraction of non-motile bacteria
\citep{Chai2011}. It is therefore of great interest to understand and characterize emergent behavior in systems composed of particles with differing motility.

\begin{figure}[b!]
\begin{tabular}{cc}
(a){\scriptsize{} $L=4,\,\phi=0.2,\,\text{Pe}=10$} & (b) {\scriptsize $L=4,\,\phi=0.8,\,\text{Pe}=120$}\tabularnewline
\includegraphics[scale=0.26]{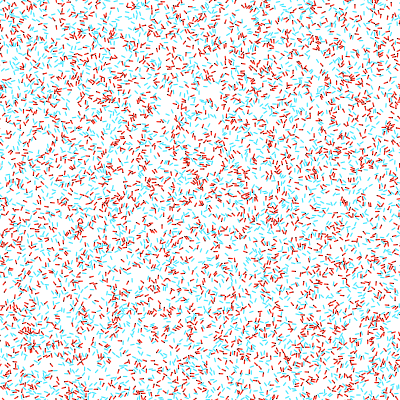} & \includegraphics[scale=0.26]{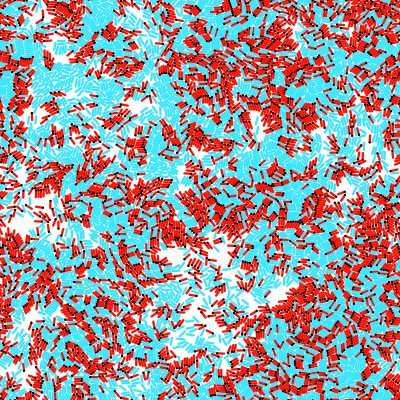}\tabularnewline
(c){\scriptsize{} $L=21,\,\phi=0.05,\,\text{Pe}=120$} & (d) {\scriptsize $L=10,\,\phi=0.8,\,\text{Pe}=120$}\tabularnewline
\includegraphics[scale=0.26]{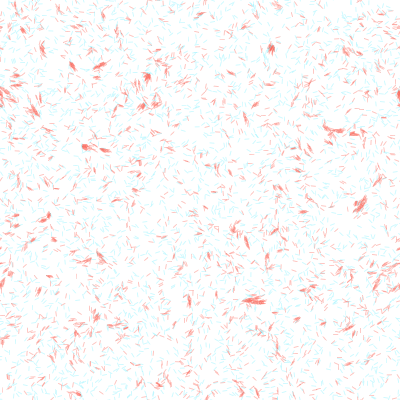} & \includegraphics[scale=0.26]{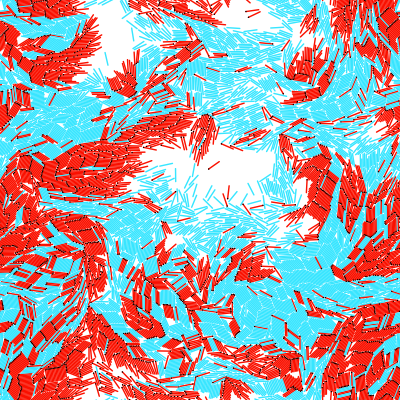}\tabularnewline
(e){\scriptsize{} $L=21,\,\phi=0.6,\,\text{Pe}=120$} & (f){\scriptsize{} $L=21,\,\phi=0.8,\,\text{Pe}=80$}\tabularnewline
\includegraphics[scale=0.26]{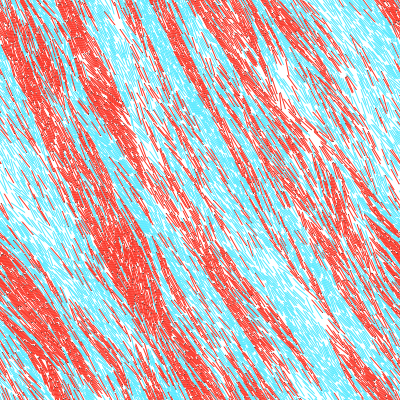} & \includegraphics[scale=0.26]{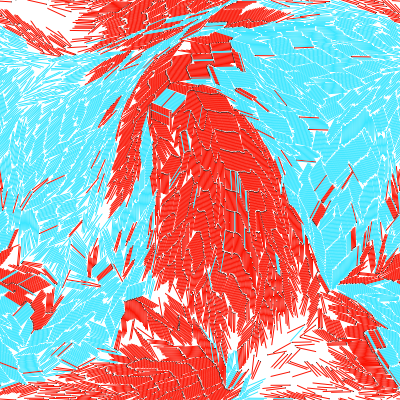}\tabularnewline
\end{tabular}

\caption{\label{fig:Example-simulations-mixed}
 Emergent structures in the mixed system: (a) isotropic mixture, (b) swarming without coherent clusters, (c) coherent clusters in passive gas,  (d) coherent active clusters with passive clustering, (e) formation of transient lanes and (f) a giant cluster after the lanes break down.
Active rods are labeled red, and passive blue; black dots indicate the head of each active rod.
$L$ is the aspect ratio, $\phi$ is the area fraction, and Pe is the Peclet Number.  Movies can be found in \cite{SupplementalInformation}.
}
\end{figure}

In this work, we computationally investigate the behavior of an example mixed motility system -- a mixture of self-propelled and passive rods interacting solely through excluded volume interactions. We show that, above threshold values of the propulsion strength and density of active rods, the mixed system is unstable to the formation of segregated domains of active and passive rods. Depending on parameters, the domain morphology and system dynamics exhibit a rich variety of behavior (Fig. \ref{fig:Example-simulations-mixed}), including uniform mixing, motile clusters of active rods in a gas of passive rods, segregated clusters of active and passive rods, and flowing bands or lanes. The degree of segregation depends on the interplay between an intrinsic tendency for particles with different collision frequencies to demix and the emergence of coherent
swarms or polar clusters that glide through a system. Our study is inspired by observations of segregation of motile and non-motile bacteria in bacterial biofilms
\citep{Chai2011}. While the factors driving bacterial segregation potentially include biological mechanisms such as chemical signaling between bacteria, recent experiments have shown that excluded volume
interactions play an important role in emergent behavior in bacterial suspensions, both in bulk and near boundaries \citep{Cisneros2011a,Li2009,Drescher2011}, and theoretical work has shown that density dependent diffusion coefficients can drive pattern formation \cite{Tailleur2008,Cates2010,Thompson2011}.  In this work, we demonstrate a generic mechanism by which excluded volume interactions and propulsion can give rise to segregation without signaling or long-range interactions.

\begin{figure}
\begin{tabular}{cc}
(a){\scriptsize{} $L=21,\,\phi=0.2,\,\text{Pe}=120$} & (b) {\scriptsize $L=10,\,\phi=0.2,\,\text{Pe}=120$}\tabularnewline
\includegraphics[scale=0.26]{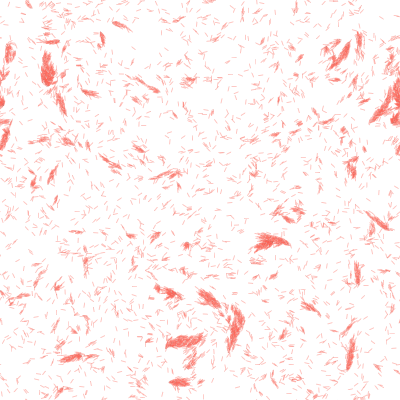} & \includegraphics[scale=0.26]{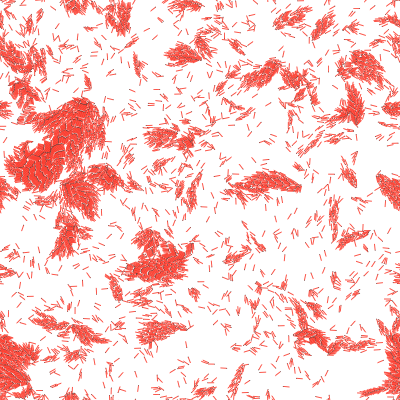}\tabularnewline
(c){\scriptsize{} $L=4,\,\phi=0.05,\,\text{Pe}=120$} & (d) {\scriptsize $L=10,\,\phi=0.8,\,\text{Pe}=120$}\tabularnewline
\includegraphics[scale=0.26]{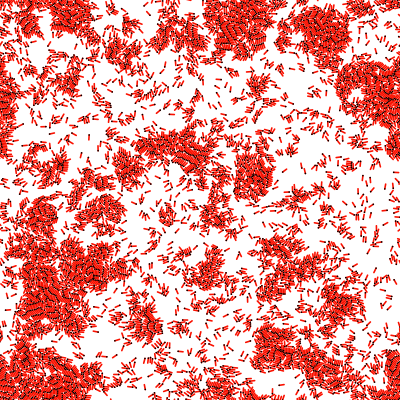} & \includegraphics[scale=0.26]{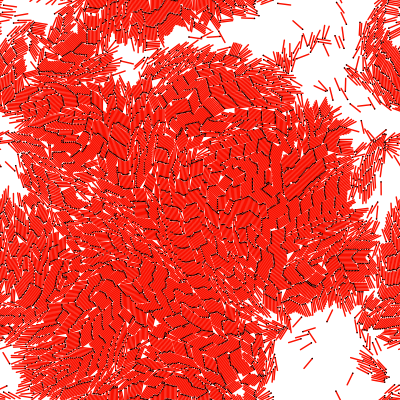}\tabularnewline
(e){\scriptsize{} $L=21,\,\phi=0.8,\,\text{Pe}=40$} & (f){\scriptsize{} $L=21,\,\phi=0.8,\,\text{Pe}=40$}\tabularnewline
\includegraphics[scale=0.26]{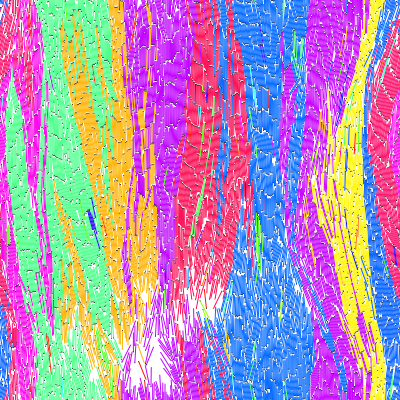} & \includegraphics[scale=0.26]{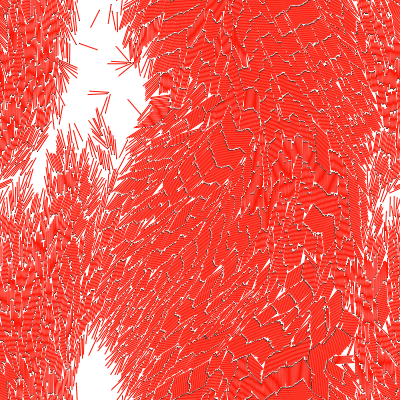}\tabularnewline
\end{tabular}

\caption{\label{fig:Example-simulations-pure}Representative snapshots from pure active systems. At low packing fraction we observe behavior consistent with that of previous studies \cite{Baskaran2008c,Bertin2006,Bertin2009,Chate2008,Chate2008,Cisneros2011a,Deseigne2010a,Ginelli2010a,Head2010a,Jia2008a,Kudrolli2008a,Narayan2007a,Peruani2006a,Peruani2011,Peruani2011a,Saintillan2011,Schaller2010a,Toner1998,Vicsek1995a,Yang2010a,Zhang2010a}: (a) small but coherent polar clusters, (b) large clusters, (c) transient apolar swarms.  New behaviors emerge at higher packing fractions: (d) a giant swarm exhibiting smectic and tetratic order, (e) lanes in a nematic prior to breakdown, (f) giant swarms appear after the lanes shown in (e) break down.
The head of each rod is indicated by a black dot, and rods within different lanes are distinguished by color in (e). The system parameters are listed above each image.  Movies can be found in \cite{SupplementalInformation}.}
\end{figure}

\section{Model}

We model rods as 2D spherocylinders (rectangles with circular
caps), represented as line segments of length $\ell$ on a two-dimensional
plane with periodic boundary conditions. We implement excluded volume
interactions between rods with a purely repulsive WCA \citep{Weeks1971a} potential $V\left(r\right)=4\kT\left[\left(\frac{r}{\sigma}\right)^{-12}-\left(\frac{r}{\sigma}\right)^{-6}\right]+\kT$,
cut off by setting $V\left(r\right)=0$ for $r>2^{1/6}$, where $r$
is the shortest distance between the two representative line segments. This potential gives the rods an effective diameter $\sigma$
and an aspect ratio of $L=\left(\ell/\sigma+1\right)$. In addition,
a fraction $\fa$ of rods are self-propelled and hence endowed
with a constant propulsive force with magnitude $F_{\text{P}}$ directed
along the rod axis.

We do not include hydrodynamic interactions in our model; these can be neglected in 2D for a dense viscous system due to screening.
We consider overdamped Brownian dynamics with the discretized equations of motion \citep{Tao2005a}
\begin{eqnarray}
\mathbf{r}\left(t+\delta t\right) & = & \mathbf{r}\left(t\right)+\mathbf{\Xi}^{-1}\left(t\right)\cdot\mathbf{F}_{\text{S}}\left(t\right)\delta t+\delta\mathbf{r}\left(t\right)\nonumber\\
\hat{\mathbf{u}}\left(t+\delta t\right) & = & \hat{\mathbf{u}}\left(t\right)+\gamma_{\text{r}}^{-1} \mathbf{T}_{\text{S}}\times\hat{\mathbf{u}}\left(t\right)\delta t+\delta\hat{\mathbf{u}}\left(t\right)
\label{eq:Equations-of-motion}
\end{eqnarray}
where $\hat{\mathbf{u}}$ is a unit vector directed along the rod
axis, $\mathbf{r}$ is the position of the rod, $\delta t$ is the
time step, $\mathbf{F}_{\text{S}}$ and $\mathbf{T}_{\text{S}}$ are
the systematic force and torque, and $\delta\mathbf{r}$ and $\delta\hat{\mathbf{u}}$
are Brownian noise terms. The friction coefficients are given by $\mathbf{\Xi}=\gamma_{\parallel}\hat{\mathbf{u}}\otimes\hat{\mathbf{u}}+\gamma_{\perp}\left(\mathbf{I}-\mathbf{\hat{u}\otimes\hat{u}}\right)$,
with $\gamma_{\perp}=2\gamma_{\parallel}$, and $\gamma_{\text{r}}=\frac{L^{2}}{6}\gamma_{\parallel}$.
The Brownian noise terms are determined by
\begin{eqnarray*}
\left\langle \delta\mathbf{r}\left(t\right)\delta\mathbf{r}\left(t\right)\right\rangle  & = & 2\kT\mathbf{\Xi}^{-1}\left(t\right)\delta t\\
\left\langle \delta\hat{\mathbf{u}}\left(t\right)\delta\hat{\mathbf{u}}\left(t\right)\right\rangle  & = & 2\kT\frac{1}{\gamma_{\text{r}}}\left(\hat{\mathbf{I}}-\hat{\mathbf{u}}\otimes\hat{\mathbf{u}}\right)\delta t.
\end{eqnarray*}
The systematic force $\mathbf{F}_{\text{S}}=\mathbf{F}_{\text{E}}+\mathbf{F}_{\text{P}}$
includes both the excluded volume interaction forces $\mathbf{F}_{\text{E}}$
and, for active rods, the uniform self propulsion force, $\mathbf{F}_{\text{P}}=F_{\text{P}}\hat{\mathbf{u}}$.

The adjustable dimensionless parameters are the aspect ratio $L$,
the propulsion force $\tilde{F}_{\text{P}}=F_{\text{P}}\left(\frac{\kT}{\sigma}\right)^{-1}$,
the area fraction $\phi$, and the motile fraction $\fa$. The Péclet
number is given by $\text{Pe}=L\tilde{F}_{\text{P}}$. We integrate Eqs.~\ref{eq:Equations-of-motion} using a second-order predictor-corrector algorithm \cite{Branka1999, Heyes2000} with an adaptive
time step with a maximum value $\delta\tilde{t}\equiv\delta t\left(\frac{\sigma^{2}\gamma_{\parallel}}{\kT}\right)^{-1}=0.0003$
\footnote{The actual timestep used for a given iteration is $\delta t/n$, where
$n$ is initially set to one. We mark an integration error if the
force on a rod is such that it will be moved more than $0.6/n$ rod
diameters, rotated more than $4\sigma/\ell n$ radians, or if the
lines representing two rods overlap. Such events are inevitable in
Brownian dynamics at high density at any timestep for a sufficiently
long run because of the Gaussian noise. If an error is recorded, we
revert back to a checkpoint saved every 10,000 timesteps and increment
$n$. We decrement $n$ after 10,000 successful timesteps.%
}.


\emph{Behavior of pure active rods:} We performed simulations of $N=6400$ rods with aspect ratios $L=4,\,10,\,21$,
for area fractions $\phi\in[0.05,0.8]$, and Péclet numbers $\text{Pe}\in\left[10,120\right]$.  Initially, we
set $\fa=1.0$ to study the behavior of the purely active system.
We initialized all systems with states equilibrated without propulsion
forces. At low Pe and $\phi$ the system remains disordered and homogeneous in density.
Increasing Pe or $\phi$ leads to aggregation
of the rods; while long rods $L=10,\,21$ form long-lived polar clusters, the short rods $L=4$ form transient aggregates which can
better be described as swarms (see Fig. \ref{fig:Example-simulations-pure}).

Clustering of self-propelled
rods has been studied extensively \cite{Baskaran2008c,Bertin2006,Bertin2009,Chate2008,Chate2008,Cisneros2011a,Deseigne2010a,Ginelli2010a,Head2010a,Jia2008a,Kudrolli2008a,Narayan2007a,Peruani2006a,Peruani2011,Peruani2011a,Saintillan2011,Schaller2010a,Toner1998,Vicsek1995a,Yang2010a,Zhang2010a} and arises because
collisions between self-propelled rods lead to alignment and correlated motion \citep{Baskaran2008c,Toner2005}. Most closely related to our work, in a study of pure self-propelled rods Yang et al. \cite{Yang2010a} find that the system exhibits three qualitatively different cluster size distributions depending on Pe. At low Pe, they observe a power law distribution with an exponential cutoff, above a threshold Pe the distribution develops a plateau at large cluster sizes (as seen in the pioneering simulations of Peruani et al. \cite{Peruani2006a}), and for high Pe they observe jammed immobile clusters. In comparison, we will focus on low and moderate values of Pe and we observe only the first two forms of cluster size distributions (a detailed description of the cluster size distributions is given in the SI, Fig.~\ref*{Supplement-fig:cluster-size-example}  \cite{SupplementalInformation}).

We examine a larger range of packing fractions than has been simulated previously \cite{Peruani2006a,Yang2010a} and we find that large packing fractions of self-propelled rods lead to some qualitatively new behaviors (Fig.~\ref{fig:Example-simulations-mixed}d-f). For example, long rods ($L=21$) above the passive isotropic nematic transition
$\phi \approx 0.22$ form multiple bands (Fig. \ref{fig:Example-simulations-mixed}e) or {}``lanes'' (e.g.
\citep{Vissers2011,Valiveti1999,Batchelor1986,Cox1990,Couzin2003,Milgram1969,Delhommelle2005,Dzubiella2002,Loewen2010,Rex2007,Sutterlin2009,Wysocki2011}). Within a band, all
rods are moving in the same direction. In contrast to human pedestrians \cite{Milgram1969} or ants \cite{Couzin2003}, the rods choose their flow directions spontaneously. The bands appear to be transient,
however, since in most of our simulations they eventually break down
when a small group of active rods begins moving perpendicular to the
direction of nematic order, shattering the nematic order and generating
configurations like that shown in Fig. \ref{fig:Example-simulations-mixed}f. The time before
breakdown decreases with increasing Pe, and once a system of lanes has broken down
we never see long-lived reappearance of nematic order
\footnote{The bands did not break down within our finite simulation time for
any of our systems with Pe = 10, but we
anticipate that breakdown would eventually occur for sufficiently
long simulations.}. Another feature of our clusters is smectic ordering, reminiscent of behavior seen in vibrated granular systems \cite{Narayan2006}, which we
attribute to the high packing fraction found in these clusters. Finally, we observe that short rods ($L=4$) form swarms which are qualitatively different from the clusters formed by long rods, with significantly less polar order and more frequent turnover of rods. In the present work we will focus on the behavior that emerges when passive and self-propelled rods are mixed; we will describe the behavior of pure self-propelled rods at high packing fractions in greater detail in a future work.

\begin{figure}
\includegraphics[width=\columnwidth]{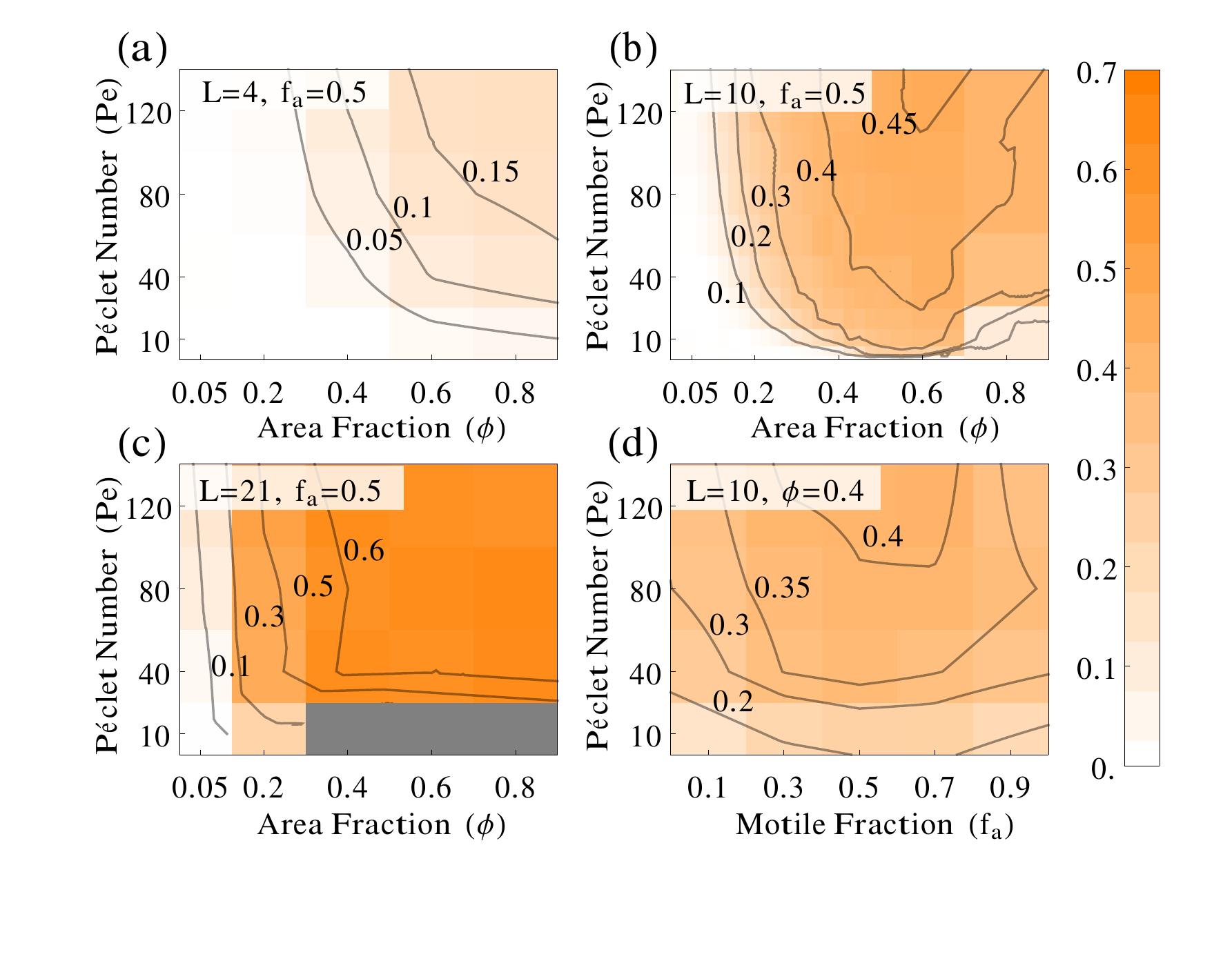}

\caption{\label{fig:Segregation-parameter}Degree of segregation $\hat\sigma$ over
(a-c) varying aspect ratio, Péclet number, and area fraction; (d) varying Péclet number and motile fraction.
Simulations which have not reached steady-state are not included in the average, and parameters with no steady-state
systems are marked in gray.  Contours are drawn using an interpolating function.}
\end{figure}

\section{Segregation in Mixed Systems}

\emph{Descriptive phenomenology:} To study the mixed system, we performed simulations with the same ranges of parameters as for the purely active systems, but with varying motile fractions $0.1\le\fa\le0.9$. Except where mentioned otherwise, we will focus on $\fa=0.5$. A variety of emergent structures appear in this range of parameters (see Fig. \ref{fig:Example-simulations-mixed}).
At low Pe and $\phi$ the system remains homogeneous in density and composition (Fig. \ref{fig:Example-simulations-mixed}a).  At higher
Pe and $\phi$, the active rods aggregate and form clusters or swarms, as in the purely active case (Fig. \ref{fig:Example-simulations-mixed}b).
Over a narrow range of Pe and $\phi$ the diffusive timescale $\tau_{\text{d}}$
of the passive rods is short in comparison to the time between collisions
with clusters, and the active clusters are surrounded by a {}``gas''
of passive rods, whose spatial properties are relatively unaffected
by the activity (Fig. \ref{fig:Example-simulations-mixed}c). Upon increasing Pe or $\phi$ collisions
become fast in comparison to $\tau_{\text{d}}$ and the passive rods
exhibit large density fluctuations (Fig. \ref{fig:Example-simulations-mixed}d). Essentially, the
forward motion of active clusters forces the passive rods into dense
{}``herds'', leaving regions in their wake nearly vacant.  Finally, for long rods
above the isotropic-nematic transition, active and passive rods initially form separate
lanes (Fig. \ref{fig:Example-simulations-mixed}e). In the system with $\fa=0.5$ each active lane is separated by a band of passive rods. Thus, in contrast to the active system where neighboring lanes flow in opposite directions, we did not observe a correlation between directions of motion of nearest active lanes. However, the active lanes do impart some extra motion to rods at the edges of passive lanes.     As in the purely active systems, these lanes are unstable and ultimately break down to yield giant clusters of active rods (Fig. \ref{fig:Example-simulations-mixed}f).

\emph{Segregation order parameter:} To quantify the dependence of the degree of segregation of the active and
passive rods on the control parameters, we employ an order parameter commonly used to measure the degree to which two human population groups
live separately from one another \citep{Massey1988}. We divide the
system into boxes of side length $\lcg$, and compute a weighted sum
over boxes
\[
\sigma_{a}\left(b\right)=\frac{1}{2f_{\text{a}}\left(1-f_{\text{a}}\right)}\sum_{i}\left(\frac{n_{i}}{n_{\text{tot}}}\right)\left|f_{i}-f_{\text{a}}\right|,
\]
where $n_{i}$ is the total number of rods in box $i$, $f_{i}$ is
the motile fraction within the box, $n_{\text{tot}}=\sum_{i}n_{i}$, and
the prefactor ensures that $\sigma_{a}\in[0,1]$ with 0 corresponding to the theoretical minimum possible segregation and 1 corresponding to maximal segregation.   It is well known that measurement of segregation is hindered by the modifiable unit area problem \cite{Massey1988} meaning that the value of the order parameter depends on box arrangement and coarse-graining length $\lcg$. For example, choosing $\lcg$ smaller than the rod diameter would yield a large value of segregation even for $\text{Pe}=0$ due to the hard core repulsions, while choosing $\lcg$ comparable to the simulation box size would result in no apparent segregation for any value of Pe.  We overcome this problem in two ways. First, we use overlapping boxes. Second, noting that the order parameter is nonzero for Péclet number zero, we subtract the
same quantity $\sigma_{0}$$\left(b\right)$ computed for a system
with the same parameter values except $\text{Pe}=0$. The difference in segregation $\sigma_{a}-\sigma_{0}$ is nonmonotonic with respect to $\lcg$ (as shown SI Fig.~\ref*{Supplement-fig:Segregation-determination}  \cite{SupplementalInformation}). We report the maximum value of this quantity over $b$, $\hat{\sigma}=\text{max}_{b}\left[\sigma_{a}\left(b\right)-\sigma_{0}\left(b\right)\right]$. Because $\hat{\sigma}$ subtracts the small degree of segregation that occurs naturally in passive systems, its maximum possible value at complete segregation is approximately 0.75 (this value depends weakly on area fraction).

\emph{Dependence of segregation on system parameters:} As shown in  Fig.~\ref{fig:Segregation-parameter}, the degree of segregation $\hat{\sigma}$ generally increases with increasing Péclet number, area fraction, and aspect ratio, with the large aspect ratio rods approaching perfect segregation for large Pe and $\phi$. However, the degree of segregation decreases mildly for large $\phi$ and small $L$ (Fig. \ref{fig:Segregation-parameter}b) where it is frustrated by jamming. 
For $L=21$ and low Pe, the nematic lanes did not break down within our simulation time (these parameter values are indicated by gray boxes in Fig.~\ref{fig:Segregation-parameter}c).
Finally, as shown in Fig.~\ref{fig:Segregation-parameter}d, the degree of segregation is nonmonotonic with respect to the motile fraction $\fa$, the maximum occurring roughly at the case of a symmetric mixture $\fa=0.5$.   Snapshots illustrating these behaviors are found in the SI, Figs.~\ref*{Supplement-fig:frames-l10-d08}-\ref*{Supplement-fig:frames-l21-a05}   \cite{SupplementalInformation}.

\begin{figure}
\includegraphics[width=\columnwidth]{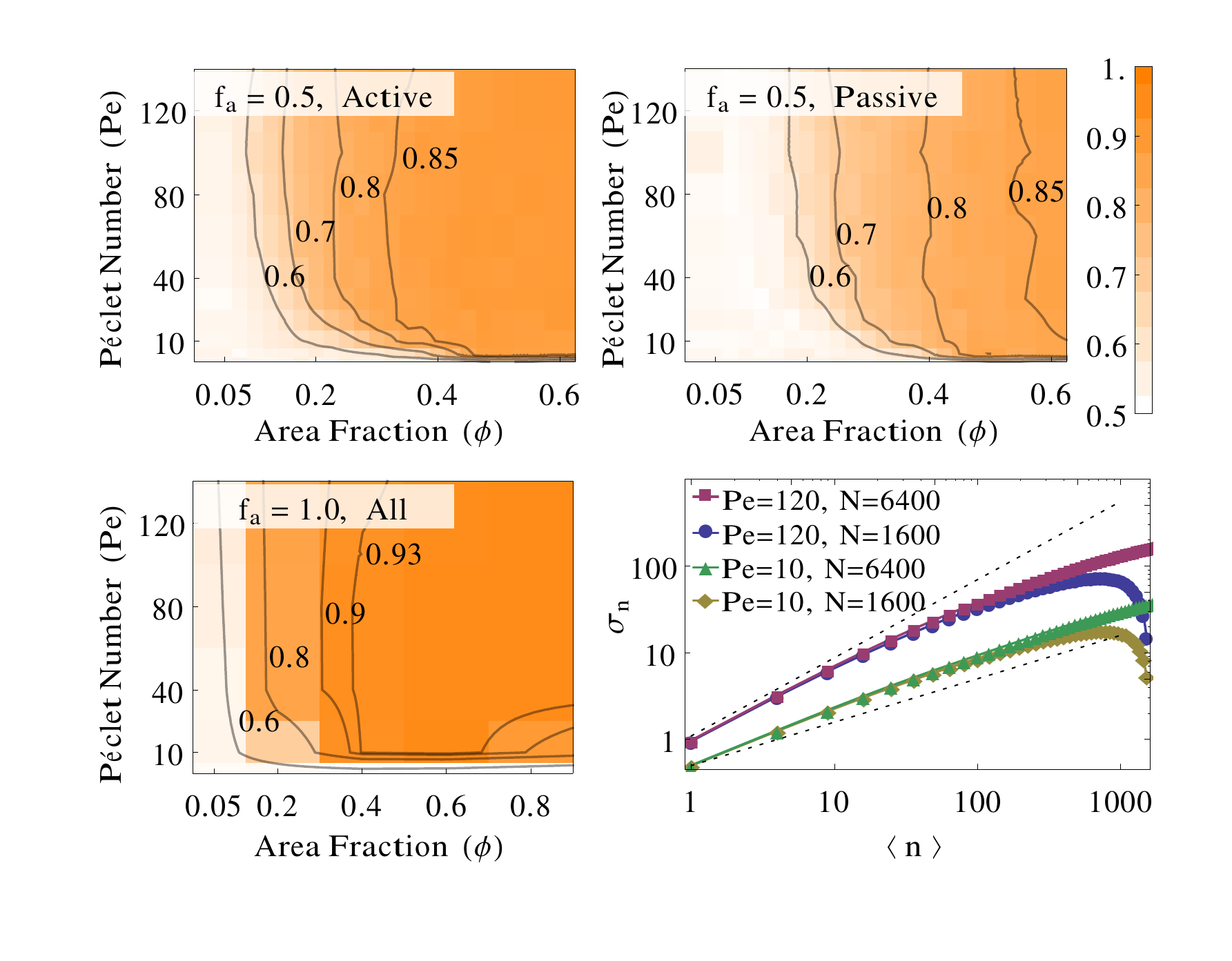}
\caption{\label{fig:Fluctuation-exponent} Density fluctuations for
active and passive rods separately in mixed systems, and for all rods in pure active systems.
The parameter is the exponent $\delta$ of the scaling relation $\sigma_{n}\propto\left\langle n\right\rangle ^{\delta}$ for the fluctuations of density in boxes which are a subset of the
entire simulation box. The exponent is calculated for regions with
mean number $1<\left\langle n\right\rangle <100$ to avoid finite
size effects, shown in the bottom figure.  Contours are drawn using an interpolating function.}
\end{figure}

\section{Anomalous Density Fluctuations}

While we have already noted that clustering in our system leads to large magnitude density fluctuations, it has been predicted theoretically \cite{Ramaswamy2003,Toner2005} and shown numerically \cite{Chate2008b,Chate2008,Chate2006} and experimentally \cite{Kudrolli2008a,Deseigne2010a,Zhang2010a,Narayan2007a} that active systems with orientational order exhibit `giant' number fluctuations (GNF), meaning that the variance grows faster than the mean as one increases the size of a region of interest. To understand the degree to which activity transfers from the self-propelled rods to their passive neighbors, we monitored number fluctuations of both active and passive rods. Specifically, we chose square regions of interest of varying size distributed throughout the system and measured separately the number of active and passive particles $n_\text{a}(t),n_\text{p}(t)$ as a function of time. We find that the scaling of number fluctuations with respect to the mean $\langle n\rangle$ exhibits a power law $\sigma \propto \langle n \rangle^\delta$ for boxes with $1<\left\langle n\right\rangle <100$; density fluctuations in larger boxes are suppressed by our finite system size of 6400 rods (Fig.~\ref{fig:Fluctuation-exponent}). As expected, the active particles exhibit GNF, meaning that $\delta>0.5$. Although previous experiments and simulations of self-propelled rods or discs found scaling exponents in the range 0.75-0.8 \cite{Deseigne2010a,Chate2008b}, we find (Fig. \ref{fig:Fluctuation-exponent}) that $\delta$ gradually increases with Pe and $\phi$, and appears to saturate at about $\delta\approx0.9$, with a sharper increase for larger aspect ratios.  Similar numbers are observed if we measure fluctuations for all rods, regardless of their motility state, and we exclude any systems which have not reached steady-state from the average.

The fact that the density fluctuation exponent undergoes a gradual transition can be understood by noting that a theoretical estimate of $\delta=1$ \cite{Ramaswamy2003,Toner2005} was derived for the case of an active nematic, where the active rods enhance density fluctuations by advecting in the direction of alignment. We expect a different result for an active but isotropic system. For parameter values at which the measured density fluctuation exponent gradually increases with Pe and $\phi$, the fraction of rods incorporated within aligned clusters is gradually increasing. Consistent with this explanation, the increase is most gradual for the aspect ratio 4 rods, which would not form a nematic phase when passive.

As shown in Fig. \ref{fig:Fluctuation-exponent}, we find that the passive rods also show GNF, with a similar scaling exponent $\delta\approx0.9$ for large activities and densities. However, the value of $\delta$ increases more slowly with activity and density than it does for the active rods. The value of the scaling exponent is correlated to the amount of clustering of passive rods; as described above, passive rods cluster extensively at large values of Pe, when they collide with active rods frequently in comparison to the diffusive timescale. In contrast, for moderate self propulsion velocities collisions are  relatively infrequent, relatively little of the activity is transferred from active rods to their passive neighbors, and thus the behavior of the passive rods resembles that of a pure equilibrium system (Fig. \ref{fig:Example-simulations-mixed}c).

\section{Origins of segregation}
\label{sec:origins}

Analysis of the segregation phase diagram (Fig.~\ref{fig:Segregation-parameter}) shows that multiple mechanisms control segregation in our system. In this section, each of these mechanisms and the interplay among them is discussed.

\emph{Collision frequency disparity:} First, due to their self propulsion, active rods experience more frequent collisions than their passive counterparts. The following simple analysis shows that this difference in collision frequencies can provide a generic mechanism for segregation. This is the dominant physics both at low Pe and low $\phi$ and at initial times at higher packing fractions.

Given an
overdamped system of two species of particles, the macroscopic description
will be a coupled set of diffusion equations for the density of active and
passive rods $\ra$ and $\rd$ $\ $respectively,%
\begin{align*}
\partial _{t}\ra=&\Daa \nabla ^{2}\ra +\Dad \nabla ^{2}\rd \\
\partial _{t}\rd=&\Ddd \nabla ^{2}\rd +\Dda \nabla ^{2}\ra
\label{eq:crossdiffusion}.
\end{align*}
Linearizing the the above equations about a homogeneous mixed state $
\rho _\text{a,p}=\rho _\text{a,p}^{0}+\delta \rho _\text{a,p}$, introducing a Fourier
transform $\widetilde{X}=\int dre^{i\mathbf{k}\cdot \mathbf{r}}X$ and
looking for solutions of the form $\widetilde{\rho }=e^{s\left( k\right) t}$%
, we can identify the linear modes of this system as
\begin{align*}
s_{\pm }\left( k\right) =& -\frac{\left( \Daa+\Ddd\right) }{2}k^{2}\\
& \pm \frac{k^{2}}{2}\sqrt{\left( \Daa-\Ddd\right) ^{2}+4\Dad\Dda}
\end{align*}%
Clearly the $+$ mode goes unstable whenever%
\begin{equation}
\Dad \Dda>\Daa \Ddd
\label{eq:LSA}.
\end{equation}%
This is the manifestation of the well-known phenomenon of cross-diffusion among species rendering the homogeneous mixed state unstable.

We can estimate the parameters in Eq.~\ref{eq:LSA} as follows. (1) To leading order in density, the diffusion of the passive rods
captured in $\Ddd$ will correspond to the ``thermal'' diffusion coefficient of an isolated rod, $\Ddd\sim \kT/\gamma \equiv D_{0}$ where $\gamma=\left( \gamma _{\parallel }+\gamma _{\bot }\right) /2$ is a mean friction coefficient.

(2) In addition to thermal diffusion, the active rods undergo
persistent motion due to their self-propulsion. This persistent motion is rendered diffusive at long times by rotational diffusion and enhances the diffusion coefficient, which can be estimated as \cite{Hagen2011} $\Daa \sim \kT/\gamma +v_{0}^{2}/D_\text{R} = D_{0}\left( 1+\text{Pe}^2\right)$ where we used $\text{Pe}\cong\frac{v_{0}^{2}}{D_{0}D_\text{R}}$.

(3) The cross diffusion terms arise from the interactions among
active and passive rods. These can be estimated from the microscopic
collisional dynamics by considering the short range repulsive interaction among the rods in
the limit of hard particle collisions. In this limit, the dynamics of the
density of one of the species, say the active rods, $\ra$, due to
collisions with the other species, in this case the passive rods, takes
the form
\begin{equation*}
\left( \partial _{t}\ra\right) _\text{coll}=-\nabla \cdot \Jad^\text{coll}.
\end{equation*}%
Here $\Jad^\text{coll}$ is the momentum transfer due to collisions between
particles of the two species and can be estimated as $\Jad^\text{coll}\sim
\frac{1}{\gamma }\Fda\ra$ with $\Fda$ the mean field force
acting on the active rods due to collisions with the passive rods. This
force is of the form  $\Fda\left( r\right) \sim \left\langle
I\sigma^\text{C}\Delta \mathbf{p}\right\rangle $ where $I$ is the incoming flux
of particles, $\sigma^\text{C}$ is the collision cross section and $\Delta
\mathbf{p}$ is the momentum transfer during such a collision with $%
\left\langle {}\right\rangle $ denoting the mean field average. The
incoming flux scales as $I\sim \rd|\mathbf{v}_{1}-\mathbf{v}_{2}|$ with $\mathbf{v}_i$ the velocity of particle $i$, and
the collision cross section $\sigma^\text{C}$ for a rod in two dimensions scales with its length $L$.
Further, for momentum conserving collisions, the momentum transfer at
contact is of the form $\left( (\mathbf{v}_{1}-\mathbf{v}_{2})\cdot \mathbf{\hat{k}}\right)\mathbf{\hat{k}} $, i.e.,
proportional to the relative velocity and directed along the normal at
the point of contact $\mathbf{\hat{k}}$. Finally, we must account for the fact that these are
collisions among particles of finite size and therefore the relevant flux $I$
in the evaluation of the mean field force should be $I\sim \rd\left( \mathbf{r}-
\bm{\zeta }\right) |\mathbf{v}_{1}-\mathbf{v}_{2}|$ where $\bm{\zeta} $ is
the contact vector, which joins the center of mass of the colliding particle to
the point of contact and typically scales as $L$.
By assuming that the velocities of the active particles have a
Maxwell-Boltzmann distribution of the form $\exp \left( -\frac{m\left(
\mathbf{v}-v_{0}\mathbf{\hat{u}}\right) ^{2}}{2\kT}\right) $ and the
velocities of the passive particles have a distribution $\exp \left( -\frac{m%
\mathbf{v}^{2}}{2\kT}\right) $, and expanding the nonlocal density in
gradients, the mean field force can be estimated to be $\Fda\left(
r\right) \sim -L^{2}\left( \kT+mv_{0}^{2}\right) \nabla \rd$. A
similar estimate can be made for the collisional contribution to the
dynamics of the density of passive rods. To transfer this analysis from the hard particle limit to our overdamped continuous system, we equate the thermal velocity $v_\text{th}=\sqrt{\kT/m}$ with the mean rate of displacement due to diffusion $\sqrt{D_0 D_\text{R}}$.
Thus, the cross
diffusion coefficients in Eq.~\ref{eq:LSA} are $\Dad\sim \frac{%
\kT}{\gamma}\left(1+\text{Pe}^{2}\right)\ra L^{2}$ and $\Dda\sim \frac{%
\kT}{\gamma}\left(1+\text{Pe}^{2}\right)\rd L^{2}$ respectively.

Substituting the estimated values of these parameters into Eq.~\ref{eq:LSA} we find that this mechanism renders the homogeneous
mixture of active and passive rods unstable whenever%
\begin{equation}
\left(1+\text{Pe}^2\right)\ra \rd L^{4} \gtrsim 1
\label{eq:threshold}
\end{equation}

\begin{figure}
\includegraphics[width=\columnwidth]{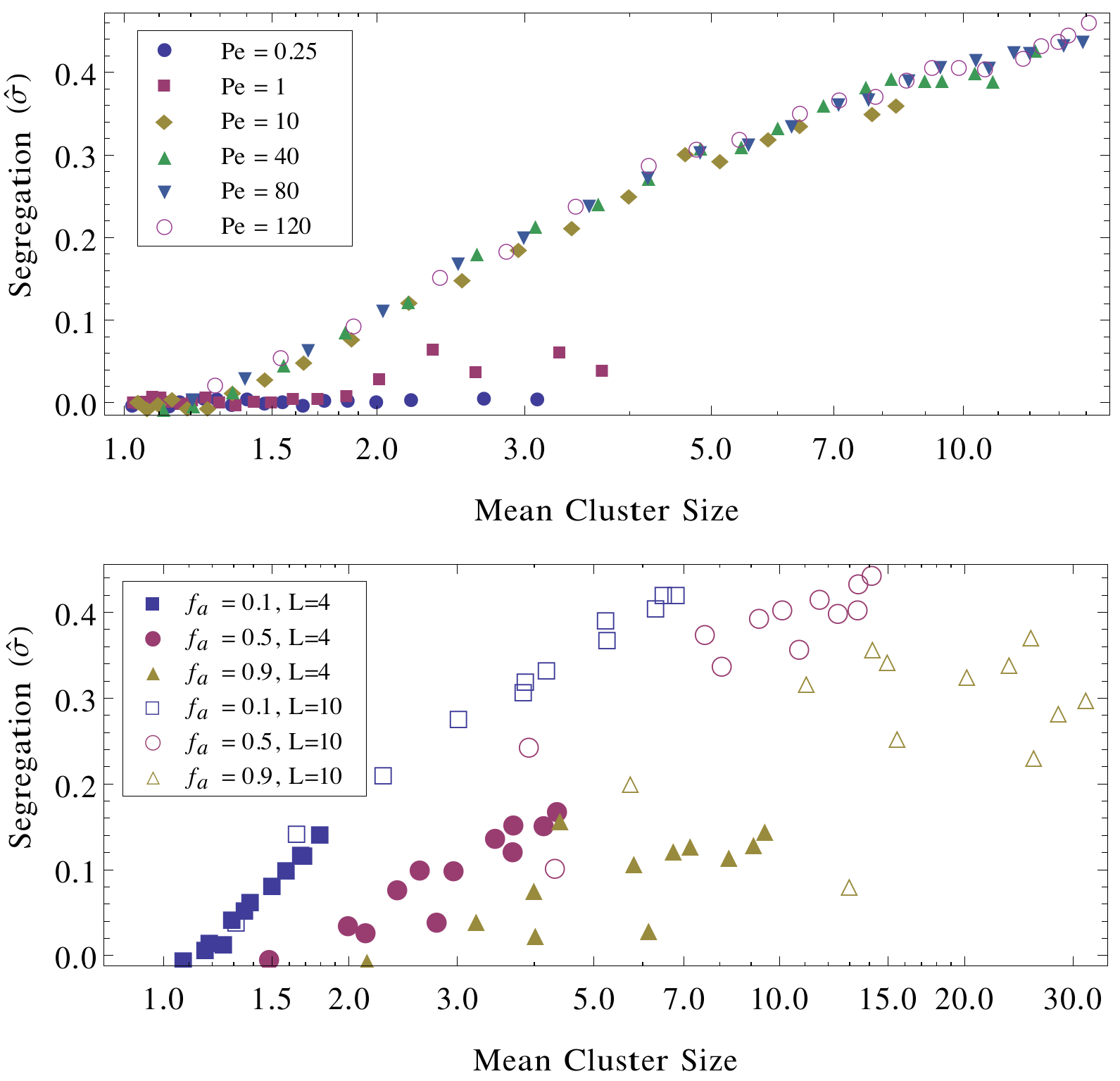}

\caption{\label{fig:Segregation-versus-clusterSize}Segregation parameter $\hat{\sigma}$
versus mean cluster size. Top: Parameters are $L=10,\,\fa=0.5$, varying $\phi\in[0.05,0.6]$ for each Pe.  The segregation $\hat\sigma$ and the mean cluster size are strongly correlated.
The points that do not fall on the main curve are due to the identification of clusters in the passive system due to random clustering.
Bottom: The relationship between mean cluster size and segregation for $\phi\in[0.05,0.8]$ and $\text{Pe}\in[10,120]$ changes with varying $\fa$, but not as markedly with varying $L$.
For $\fa$ close to 1.0, the relationship weakens because segregation is no longer governed solely by the formation of coherent clusters.}
\end{figure}

\begin{figure}
\includegraphics[width=\columnwidth]{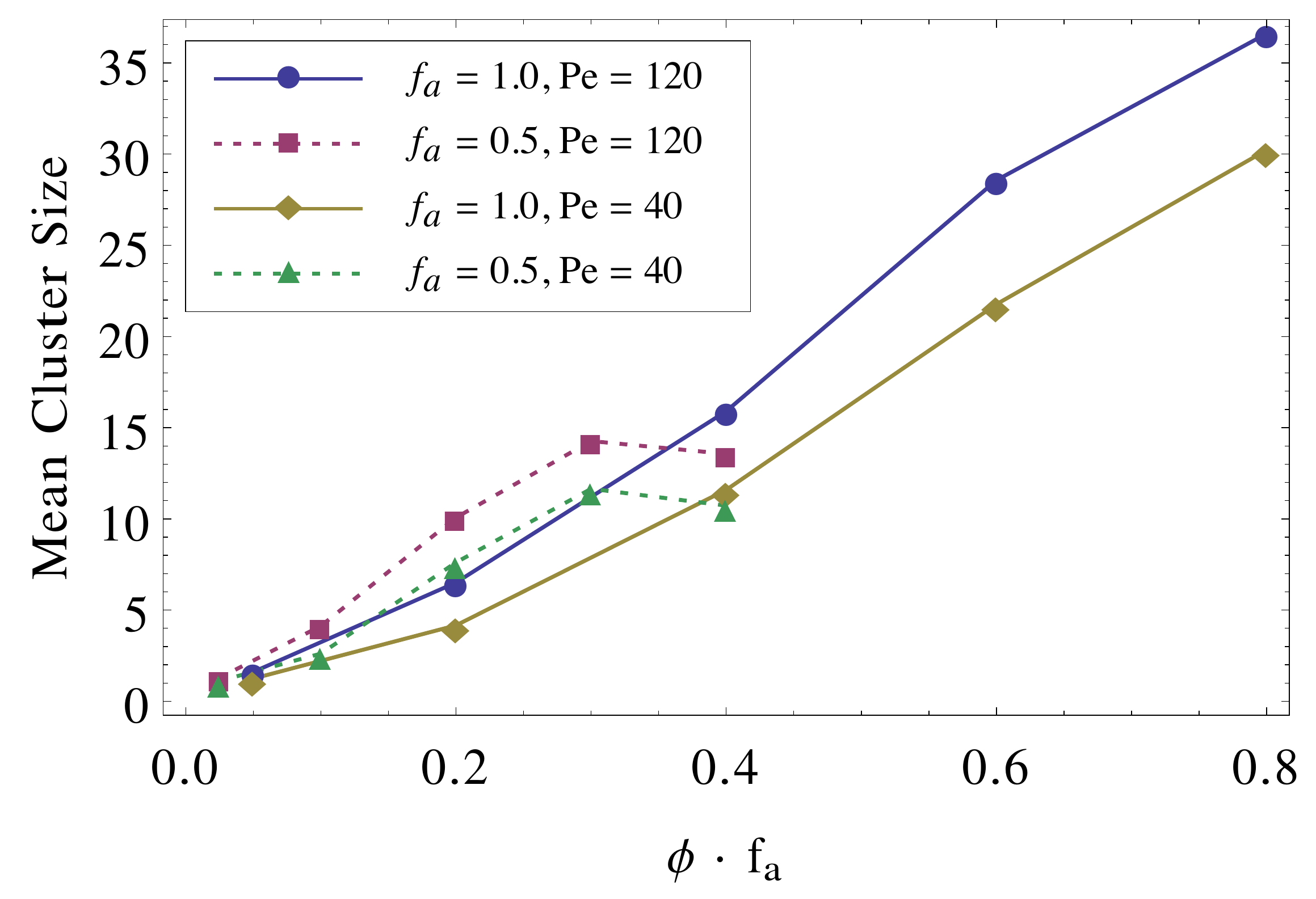}
\caption{\label{fig:meanClusterSize-comparison} Comparison of mean cluster sizes between purely active and mixed systems.  Holding the
density of active rods fixed and adding passive rods enhances clustering for $\phi \cdot \fa$ less than about 0.3, but weakens it at very high area fractions by limiting motion.
Parameters are $L=10$, varying $\phi\in[0.05,0.8]$ for each $\fa$.}
\end{figure}

It is important to note that the preceding calculation neglects numerical factors which are not necessarily of order 1, and can be expected to break down once clustering occurs since we have not accounted for polar order. However, it does illustrate the mechanism by which disparities in collision frequencies can lead to segregation, and we expect the trends predicted in Eq.~\ref{eq:threshold} to be relevant. Indeed, the prediction that segregation occurs more readily with increasing Pe, packing fraction and aspect ratio is consistent with all of the simulation results at applicable parameter values. Finally, we note that analogous effects lead to clustering and/or lane formation in colloidal systems for which interaction with external fields leads to different particle mobilities (e.g. sedimentation of particles with different masses \cite{Batchelor1986,Cox1990,Valiveti1999}) or electrophoresis of particles with different charges (e.g. \cite{Vissers2011}).

\emph{Positive feedback between clustering and segregation:} For much of the parameter space, the generic collision induced mechanism for segregation appears to be enhanced by the clustering that occurs in our self-propelled rod system. We find that the segregation behavior is closely correlated with the clustering phenomenon associated with self-propelled rods (see Fig.~\ref{fig:Segregation-versus-clusterSize}). At the same time, segregation also induces further clustering.
The alignment resulting from collisions of self-propelled particles can be thought of as an effective attraction between active particles, limiting diffusion in the transverse direction \citep{Baskaran2008c}. Since this effect is smaller for active/passive collisions, it generates an additional driving force for segregation. The feedback between clustering and collision-induced segregation is reminiscent of the mechanism by which density-dependent diffusion coefficients can amplify density inhomogeneities to provide a route to phase separation in bacterial suspensions \cite{Tailleur2008,Cates2010,Thompson2011}, but has a different origin. While the latter effect arises because bacterial motility is reduced in dense regions, the dominant effect driving segregation here is the suppression of the transverse diffusion coefficient due to alignment.  The fact that diffusion is impeded at high densities actually inhibits segregation, as evidenced by the reduction in the segregation parameter for high $\phi$ in Fig.~\ref{fig:Segregation-parameter}b.
 
 As shown in Fig.~\ref{fig:Segregation-versus-clusterSize}, there is a close relationship between the extent of clustering, measured by the mean cluster size, and the segregation parameter $\hat\sigma$ at moderate motility fractions $\fa\le0.5$, with values of $\hat{\sigma}$ from simulations at different Pe, $\phi$, and $L$ collapsing onto a universal curve for $\text{Pe}>3$.
(For $\text{Pe} < 1$ the `clusters' identified by our clustering algorithm correspond to active rods which transiently colocalize as in a passive system.)
We show in the SI (Fig.~\ref*{Supplement-fig:Segregation-vs-interiorFraction}  \cite{SupplementalInformation}) that this relationship can be understood by calculating the fraction of active rods which are in the interiors of clusters, based on the observations that most active rods are in well-defined active clusters and active clusters are `pure', meaning that they incorporate only a small fraction (of order 1\%) of passive rods. The cluster purity depends on Pe but is roughly independent of the overall motility fraction.

It is important to emphasize that a tendency to cluster alone is not sufficient to explain the data in Fig.~\ref{fig:Segregation-versus-clusterSize}. There must be a mechanism driving passive rods out of active clusters and vice versa in order for the geometric relationship to hold.  Indeed, the relationship does break down for high motility fractions (see Fig.~\ref{fig:Segregation-versus-clusterSize}b), where clusters are less well-defined, a significant fraction of the passive rods are trapped within active clusters, and herds of passive rods are transiently associated with active clusters by the mixing mechanism described below.

While the preceding evidence establishes that clustering enhances segregation, comparison between the pure active ($\fa=1$) and mixed ($\fa=0.5$) cases indicates that segregation also amplifies clustering at moderate packing fractions.  This effect can be seen in  Fig.~\ref{fig:meanClusterSize-comparison}, where mean cluster sizes are shown as a function of the density of active rods, $\phi\fa$, for $\fa=1$ and $\fa=0.5$. We see that introducing passive rods ($\fa=0.5$) yields larger clusters than do pure active simulations.  We attribute the increase in cluster size to the fact that, once segregation occurs, the active rods experience an effectively higher density which enhances their frequency of interactions.  Equivalently, the passive rods limit the transverse diffusivity of the active rods, which facilitates the formation of clusters. In contrast, at higher overall area fractions the introduction of passive rods \emph{decreases} clustering. Passive rods inhibit clustering of active rods at high densities by hindering their propulsive motion and alignment.

\emph{Cluster-collision induced mixing:} A final mechanism affecting segregation is mixing due to collisions of active clusters. As described earlier, active clusters tend to drive locally dense regions of passive rods in front of them. During cluster collisions these passive rods can be trapped within clusters. Hence, frequent cluster collisions lead to a decrease in segregation and thus increased clustering under these conditions can inhibit segregation.

The three mechanisms just described (collision frequency disparities, clustering, mixing) lead to a rich  interplay among the motions of active and passive rods, segregation, and clustering that results in a nonmonotonic variation of the degree of segregation with the motility fraction. At high motility fractions, the  small number of passive rods in the system are incorporated within clusters, leading to slightly lower magnitudes of segregation at high $\fa$ (Fig.~\ref{fig:Segregation-parameter}d). As the mean composition of passive rods ($1-\fa$) increases beyond this fraction, the generic segregation mechanism causes additional passive rods to be expelled from the clusters. These rods are driven into clusters of their own as described above (Fig.~\ref{fig:Example-simulations-mixed}d). Thus decreasing the motile fractions toward $\fa=0.5$ increases segregation. At low motile fractions, on the other hand, the preponderance of passive rods limits interactions between active rods and clustering, hence inhibiting segregation by either mechanism.

\section{Conclusion}

In summary, we have characterized mixtures of self-propelled and passive rods. We find that activity drives segregation under a wide range of system densities, propulsion velocities and compositions. By studying a minimal model, we show that segregation does not require processes such as chemotaxis, inter-organismal signaling, or hydrodynamic interactions, but rather can arise generically in any system in which particles exhibit variable motility. The alignment effects of the self-propulsion and rod geometry of the particles studied here enhance segregation, as revealed by the dependence of segregation on aspect ratio.

Because the segregation mechanism found here arises from a minimal model, it can be studied in a wide variety of experimental systems. For example,  our results suggest that motile bacteria mixed with bacteria  whose motility is genetically disabled should segregate in a range of densities and swimming speeds, although the model here would only apply for a dense two-dimensional system since we have not considered hydrodynamics. The predictions could be more directly compared to systems of propelled synthetic rods. Shaken granular rods \cite{Kudrolli2008a,Narayan2007a,Narayan2006,Prevost2002,Blair2003,Aranson2007,Galanis2006,Daniels2011},
for example, could be examined with a mixture of polar and apolar
rods, and rods propelled by catalytic reactions \cite{Paxton2004,Palacci2010,Hong2007,Ozin2010,Thakur2011,Mino2011} could be mixed with passive rods on a substrate. We anticipate that understanding the relationship between differences in activity of subpopulations and their segregation within these model systems could have implications for systems such as biofilms \cite{Chai2011,Cogan2011}, where these generic physical mechanisms underlie complex biological interactions.

\begin{acknowledgments}
We gratefully acknowledge support from the NSF Brandeis MRSEC DMR-0820492. We are indebted to Rebecca Christianson for inspiring this line of exploration, and we thank
Luca Giomi, Oren Elrad, and Gabriel Redner for helpful suggestions.  Computational support was provided by Brandeis HPC.
\end{acknowledgments}
\bibliographystyle{apsrev4-1}
\bibliography{references}

\begin{thebibliography}{71}%
\makeatletter
\providecommand \@ifxundefined [1]{%
 \@ifx{#1\undefined}
}%
\providecommand \@ifnum [1]{%
 \ifnum #1\expandafter \@firstoftwo
 \else \expandafter \@secondoftwo
 \fi
}%
\providecommand \@ifx [1]{%
 \ifx #1\expandafter \@firstoftwo
 \else \expandafter \@secondoftwo
 \fi
}%
\providecommand \natexlab [1]{#1}%
\providecommand \enquote  [1]{``#1''}%
\providecommand \bibnamefont  [1]{#1}%
\providecommand \bibfnamefont [1]{#1}%
\providecommand \citenamefont [1]{#1}%
\providecommand \href@noop [0]{\@secondoftwo}%
\providecommand \href [0]{\begingroup \@sanitize@url \@href}%
\providecommand \@href[1]{\@@startlink{#1}\@@href}%
\providecommand \@@href[1]{\endgroup#1\@@endlink}%
\providecommand \@sanitize@url [0]{\catcode `\\12\catcode `\$12\catcode
  `\&12\catcode `\#12\catcode `\^12\catcode `\_12\catcode `\%12\relax}%
\providecommand \@@startlink[1]{}%
\providecommand \@@endlink[0]{}%
\providecommand \url  [0]{\begingroup\@sanitize@url \@url }%
\providecommand \@url [1]{\endgroup\@href {#1}{\urlprefix }}%
\providecommand \urlprefix  [0]{URL }%
\providecommand \Eprint [0]{\href }%
\providecommand \doibase [0]{http://dx.doi.org/}%
\providecommand \selectlanguage [0]{\@gobble}%
\providecommand \bibinfo  [0]{\@secondoftwo}%
\providecommand \bibfield  [0]{\@secondoftwo}%
\providecommand \translation [1]{[#1]}%
\providecommand \BibitemOpen [0]{}%
\providecommand \bibitemStop [0]{}%
\providecommand \bibitemNoStop [0]{.\EOS\space}%
\providecommand \EOS [0]{\spacefactor3000\relax}%
\providecommand \BibitemShut  [1]{\csname bibitem#1\endcsname}%
\let\auto@bib@innerbib\@empty
\bibitem [{\citenamefont {Julicher}\ \emph {et~al.}(2007)\citenamefont
  {Julicher}, \citenamefont {Kruse}, \citenamefont {Prost},\ and\ \citenamefont
  {Joanny}}]{Julicher2007a}%
  \BibitemOpen
  \bibfield  {author} {\bibinfo {author} {\bibfnamefont {F.}~\bibnamefont
  {Julicher}}, \bibinfo {author} {\bibfnamefont {K.}~\bibnamefont {Kruse}},
  \bibinfo {author} {\bibfnamefont {J.}~\bibnamefont {Prost}}, \ and\ \bibinfo
  {author} {\bibfnamefont {J.}~\bibnamefont {Joanny}},\ }\href {\doibase
  10.1016/j.physrep.2007.02.018} {\bibfield  {journal} {\bibinfo  {journal}
  {Phys. Rep.}\ }\textbf {\bibinfo {volume} {449}},\ \bibinfo {pages} {3}
  (\bibinfo {year} {2007})}\BibitemShut {NoStop}%
\bibitem [{\citenamefont {Koenderink}\ \emph {et~al.}(2009)\citenamefont
  {Koenderink}, \citenamefont {Dogic}, \citenamefont {Nakamura}, \citenamefont
  {Bendix}, \citenamefont {MacKintosh}, \citenamefont {Hartwig}, \citenamefont
  {Stossel},\ and\ \citenamefont {Weitz}}]{Koenderink2009a}%
  \BibitemOpen
  \bibfield  {author} {\bibinfo {author} {\bibfnamefont {G.~H.}\ \bibnamefont
  {Koenderink}}, \bibinfo {author} {\bibfnamefont {Z.}~\bibnamefont {Dogic}},
  \bibinfo {author} {\bibfnamefont {F.}~\bibnamefont {Nakamura}}, \bibinfo
  {author} {\bibfnamefont {P.~M.}\ \bibnamefont {Bendix}}, \bibinfo {author}
  {\bibfnamefont {F.~C.}\ \bibnamefont {MacKintosh}}, \bibinfo {author}
  {\bibfnamefont {J.~H.}\ \bibnamefont {Hartwig}}, \bibinfo {author}
  {\bibfnamefont {T.~P.}\ \bibnamefont {Stossel}}, \ and\ \bibinfo {author}
  {\bibfnamefont {D.~a.}\ \bibnamefont {Weitz}},\ }\href {\doibase
  10.1073/pnas.0903974106} {\bibfield  {journal} {\bibinfo  {journal} {Proc.
  Natl. Acad. Sci. U. S. A.}\ }\textbf {\bibinfo {volume} {106}},\ \bibinfo
  {pages} {15192} (\bibinfo {year} {2009})}\BibitemShut {NoStop}%
\bibitem [{\citenamefont {Liu}\ \emph {et~al.}(2008)\citenamefont {Liu},
  \citenamefont {Richmond}, \citenamefont {Maibaum}, \citenamefont {Pronk},
  \citenamefont {Geissler},\ and\ \citenamefont {Fletcher}}]{Liu2008b}%
  \BibitemOpen
  \bibfield  {author} {\bibinfo {author} {\bibfnamefont {A.~P.}\ \bibnamefont
  {Liu}}, \bibinfo {author} {\bibfnamefont {D.~L.}\ \bibnamefont {Richmond}},
  \bibinfo {author} {\bibfnamefont {L.}~\bibnamefont {Maibaum}}, \bibinfo
  {author} {\bibfnamefont {S.}~\bibnamefont {Pronk}}, \bibinfo {author}
  {\bibfnamefont {P.~L.}\ \bibnamefont {Geissler}}, \ and\ \bibinfo {author}
  {\bibfnamefont {D.~a.}\ \bibnamefont {Fletcher}},\ }\href {\doibase
  10.1038/nphys1071} {\bibfield  {journal} {\bibinfo  {journal} {Nat. Phys.}\
  }\textbf {\bibinfo {volume} {4}},\ \bibinfo {pages} {789} (\bibinfo {year}
  {2008})}\BibitemShut {NoStop}%
\bibitem [{\citenamefont {Mizuno}\ \emph {et~al.}(2007)\citenamefont {Mizuno},
  \citenamefont {Tardin}, \citenamefont {Schmidt},\ and\ \citenamefont
  {Mackintosh}}]{Mizuno2007a}%
  \BibitemOpen
  \bibfield  {author} {\bibinfo {author} {\bibfnamefont {D.}~\bibnamefont
  {Mizuno}}, \bibinfo {author} {\bibfnamefont {C.}~\bibnamefont {Tardin}},
  \bibinfo {author} {\bibfnamefont {C.~F.}\ \bibnamefont {Schmidt}}, \ and\
  \bibinfo {author} {\bibfnamefont {F.~C.}\ \bibnamefont {Mackintosh}},\ }\href
  {\doibase 10.1126/science.1134404} {\bibfield  {journal} {\bibinfo  {journal}
  {Science}\ }\textbf {\bibinfo {volume} {315}},\ \bibinfo {pages} {370}
  (\bibinfo {year} {2007})}\BibitemShut {NoStop}%
\bibitem [{\citenamefont {Schaller}\ \emph {et~al.}(2010)\citenamefont
  {Schaller}, \citenamefont {Weber}, \citenamefont {Semmrich}, \citenamefont
  {Frey},\ and\ \citenamefont {Bausch}}]{Schaller2010a}%
  \BibitemOpen
  \bibfield  {author} {\bibinfo {author} {\bibfnamefont {V.}~\bibnamefont
  {Schaller}}, \bibinfo {author} {\bibfnamefont {C.}~\bibnamefont {Weber}},
  \bibinfo {author} {\bibfnamefont {C.}~\bibnamefont {Semmrich}}, \bibinfo
  {author} {\bibfnamefont {E.}~\bibnamefont {Frey}}, \ and\ \bibinfo {author}
  {\bibfnamefont {A.~R.}\ \bibnamefont {Bausch}},\ }\href {\doibase
  10.1038/nature09312} {\bibfield  {journal} {\bibinfo  {journal} {Nature}\
  }\textbf {\bibinfo {volume} {467}},\ \bibinfo {pages} {73} (\bibinfo {year}
  {2010})}\BibitemShut {NoStop}%
\bibitem [{\citenamefont {Swaminathan}\ \emph {et~al.}(2008)\citenamefont
  {Swaminathan}, \citenamefont {Karpeev},\ and\ \citenamefont
  {Aranson}}]{Swaminathan2008a}%
  \BibitemOpen
  \bibfield  {author} {\bibinfo {author} {\bibfnamefont {S.}~\bibnamefont
  {Swaminathan}}, \bibinfo {author} {\bibfnamefont {D.}~\bibnamefont
  {Karpeev}}, \ and\ \bibinfo {author} {\bibfnamefont {I.}~\bibnamefont
  {Aranson}},\ }\href {\doibase 10.1103/PhysRevE.77.066206} {\bibfield
  {journal} {\bibinfo  {journal} {Phys. Rev. E}\ }\textbf {\bibinfo {volume}
  {77}},\ \bibinfo {pages} {1} (\bibinfo {year} {2008})}\BibitemShut {NoStop}%
\bibitem [{\citenamefont {Toner}\ and\ \citenamefont {Tu}(1998)}]{Toner1998}%
  \BibitemOpen
  \bibfield  {author} {\bibinfo {author} {\bibfnamefont {J.}~\bibnamefont
  {Toner}}\ and\ \bibinfo {author} {\bibfnamefont {Y.}~\bibnamefont {Tu}},\
  }\href {\doibase 10.1103/PhysRevE.58.4828} {\bibfield  {journal} {\bibinfo
  {journal} {Phys. Rev. E}\ }\textbf {\bibinfo {volume} {58}},\ \bibinfo
  {pages} {4828} (\bibinfo {year} {1998})}\BibitemShut {NoStop}%
\bibitem [{\citenamefont {Zhang}\ \emph {et~al.}(2010)\citenamefont {Zhang},
  \citenamefont {Be'er}, \citenamefont {Florin},\ and\ \citenamefont
  {Swinney}}]{Zhang2010a}%
  \BibitemOpen
  \bibfield  {author} {\bibinfo {author} {\bibfnamefont {H.~P.}\ \bibnamefont
  {Zhang}}, \bibinfo {author} {\bibfnamefont {A.}~\bibnamefont {Be'er}},
  \bibinfo {author} {\bibfnamefont {E.-L.}\ \bibnamefont {Florin}}, \ and\
  \bibinfo {author} {\bibfnamefont {H.~L.}\ \bibnamefont {Swinney}},\ }\href
  {\doibase 10.1073/pnas.1001651107} {\bibfield  {journal} {\bibinfo  {journal}
  {Proc. Natl. Acad. Sci. U. S. A.}\ }\textbf {\bibinfo {volume} {107}},\
  \bibinfo {pages} {13626} (\bibinfo {year} {2010})}\BibitemShut {NoStop}%
\bibitem [{\citenamefont {Ramaswamy}\ \emph {et~al.}(2003)\citenamefont
  {Ramaswamy}, \citenamefont {Simha},\ and\ \citenamefont
  {Toner}}]{Ramaswamy2003}%
  \BibitemOpen
  \bibfield  {author} {\bibinfo {author} {\bibfnamefont {S.}~\bibnamefont
  {Ramaswamy}}, \bibinfo {author} {\bibfnamefont {R.~A.}\ \bibnamefont
  {Simha}}, \ and\ \bibinfo {author} {\bibfnamefont {J.}~\bibnamefont
  {Toner}},\ }\href {\doibase 10.1209/epl/i2003-00346-7} {\bibfield  {journal}
  {\bibinfo  {journal} {Europhys. Lett.}\ }\textbf {\bibinfo {volume} {62}},\
  \bibinfo {pages} {196} (\bibinfo {year} {2003})}\BibitemShut {NoStop}%
\bibitem [{\citenamefont {Chat\'{e}}\ \emph
  {et~al.}(2008{\natexlab{a}})\citenamefont {Chat\'{e}}, \citenamefont
  {Ginelli}, \citenamefont {Gr\'{e}goire}, \citenamefont {Peruani},\ and\
  \citenamefont {Raynaud}}]{Chate2008}%
  \BibitemOpen
  \bibfield  {author} {\bibinfo {author} {\bibfnamefont {H.}~\bibnamefont
  {Chat\'{e}}}, \bibinfo {author} {\bibfnamefont {F.}~\bibnamefont {Ginelli}},
  \bibinfo {author} {\bibfnamefont {G.}~\bibnamefont {Gr\'{e}goire}}, \bibinfo
  {author} {\bibfnamefont {F.}~\bibnamefont {Peruani}}, \ and\ \bibinfo
  {author} {\bibfnamefont {F.}~\bibnamefont {Raynaud}},\ }\href {\doibase
  10.1140/epjb/e2008-00275-9} {\bibfield  {journal} {\bibinfo  {journal} {Euro.
  Phys. J. B}\ }\textbf {\bibinfo {volume} {64}},\ \bibinfo {pages} {451}
  (\bibinfo {year} {2008}{\natexlab{a}})}\BibitemShut {NoStop}%
\bibitem [{\citenamefont {Deseigne}\ \emph {et~al.}(2010)\citenamefont
  {Deseigne}, \citenamefont {Dauchot},\ and\ \citenamefont
  {Chat\'{e}}}]{Deseigne2010a}%
  \BibitemOpen
  \bibfield  {author} {\bibinfo {author} {\bibfnamefont {J.}~\bibnamefont
  {Deseigne}}, \bibinfo {author} {\bibfnamefont {O.}~\bibnamefont {Dauchot}}, \
  and\ \bibinfo {author} {\bibfnamefont {H.}~\bibnamefont {Chat\'{e}}},\ }\href
  {\doibase 10.1103/PhysRevLett.105.098001} {\bibfield  {journal} {\bibinfo
  {journal} {Phys. Rev. Lett.}\ }\textbf {\bibinfo {volume} {105}},\ \bibinfo
  {pages} {1} (\bibinfo {year} {2010})}\BibitemShut {NoStop}%
\bibitem [{\citenamefont {Toner}\ \emph {et~al.}(2005)\citenamefont {Toner},
  \citenamefont {Tu},\ and\ \citenamefont {Ramaswamy}}]{Toner2005}%
  \BibitemOpen
  \bibfield  {author} {\bibinfo {author} {\bibfnamefont {J.}~\bibnamefont
  {Toner}}, \bibinfo {author} {\bibfnamefont {Y.}~\bibnamefont {Tu}}, \ and\
  \bibinfo {author} {\bibfnamefont {S.}~\bibnamefont {Ramaswamy}},\ }\href
  {\doibase 10.1016/j.aop.2005.04.011} {\bibfield  {journal} {\bibinfo
  {journal} {Ann. Phys.}\ }\textbf {\bibinfo {volume} {318}},\ \bibinfo {pages}
  {170} (\bibinfo {year} {2005})}\BibitemShut {NoStop}%
\bibitem [{\citenamefont {Narayan}\ \emph {et~al.}(2007)\citenamefont
  {Narayan}, \citenamefont {Ramaswamy},\ and\ \citenamefont
  {Menon}}]{Narayan2007a}%
  \BibitemOpen
  \bibfield  {author} {\bibinfo {author} {\bibfnamefont {V.}~\bibnamefont
  {Narayan}}, \bibinfo {author} {\bibfnamefont {S.}~\bibnamefont {Ramaswamy}},
  \ and\ \bibinfo {author} {\bibfnamefont {N.}~\bibnamefont {Menon}},\ }\href
  {\doibase 10.1126/science.1140414} {\bibfield  {journal} {\bibinfo  {journal}
  {Science}\ }\textbf {\bibinfo {volume} {317}},\ \bibinfo {pages} {105}
  (\bibinfo {year} {2007})}\BibitemShut {NoStop}%
\bibitem [{\citenamefont {Chat\'{e}}\ \emph {et~al.}(2006)\citenamefont
  {Chat\'{e}}, \citenamefont {Ginelli},\ and\ \citenamefont
  {Montagne}}]{Chate2006}%
  \BibitemOpen
  \bibfield  {author} {\bibinfo {author} {\bibfnamefont {H.}~\bibnamefont
  {Chat\'{e}}}, \bibinfo {author} {\bibfnamefont {F.}~\bibnamefont {Ginelli}},
  \ and\ \bibinfo {author} {\bibfnamefont {R.}~\bibnamefont {Montagne}},\
  }\href {\doibase 10.1103/PhysRevLett.96.180602} {\bibfield  {journal}
  {\bibinfo  {journal} {Phys. Rev. Lett.}\ }\textbf {\bibinfo {volume} {96}},\
  \bibinfo {pages} {1} (\bibinfo {year} {2006})}\BibitemShut {NoStop}%
\bibitem [{\citenamefont {Kudrolli}\ \emph {et~al.}(2008)\citenamefont
  {Kudrolli}, \citenamefont {Lumay}, \citenamefont {Volfson},\ and\
  \citenamefont {Tsimring}}]{Kudrolli2008a}%
  \BibitemOpen
  \bibfield  {author} {\bibinfo {author} {\bibfnamefont {A.}~\bibnamefont
  {Kudrolli}}, \bibinfo {author} {\bibfnamefont {G.}~\bibnamefont {Lumay}},
  \bibinfo {author} {\bibfnamefont {D.}~\bibnamefont {Volfson}}, \ and\
  \bibinfo {author} {\bibfnamefont {L.}~\bibnamefont {Tsimring}},\ }\href
  {\doibase 10.1103/PhysRevLett.100.058001} {\bibfield  {journal} {\bibinfo
  {journal} {Phys. Rev. Lett.}\ }\textbf {\bibinfo {volume} {100}},\ \bibinfo
  {pages} {2} (\bibinfo {year} {2008})}\BibitemShut {NoStop}%
\bibitem [{\citenamefont {Giomi}\ and\ \citenamefont
  {Marchetti}(2011)}]{Giomi2011}%
  \BibitemOpen
  \bibfield  {author} {\bibinfo {author} {\bibfnamefont {L.}~\bibnamefont
  {Giomi}}\ and\ \bibinfo {author} {\bibfnamefont {M.~C.}\ \bibnamefont
  {Marchetti}},\ }\href {http://arxiv.org/abs/1106.1624} {\bibfield  {journal}
  {\bibinfo  {journal} {Arxiv preprint}\ ,\ \bibinfo {pages} {12}} (\bibinfo
  {year} {2011})},\ \Eprint {http://arxiv.org/abs/1106.1624} {arXiv:1106.1624}
  \BibitemShut {NoStop}%
\bibitem [{\citenamefont {Giomi}\ \emph {et~al.}(2008)\citenamefont {Giomi},
  \citenamefont {Marchetti},\ and\ \citenamefont {Liverpool}}]{Giomi2008}%
  \BibitemOpen
  \bibfield  {author} {\bibinfo {author} {\bibfnamefont {L.}~\bibnamefont
  {Giomi}}, \bibinfo {author} {\bibfnamefont {M.}~\bibnamefont {Marchetti}}, \
  and\ \bibinfo {author} {\bibfnamefont {T.}~\bibnamefont {Liverpool}},\ }\href
  {\doibase 10.1103/PhysRevLett.101.198101} {\bibfield  {journal} {\bibinfo
  {journal} {Phys. Rev. Lett.}\ }\textbf {\bibinfo {volume} {101}},\ \bibinfo
  {pages} {13} (\bibinfo {year} {2008})}\BibitemShut {NoStop}%
\bibitem [{\citenamefont {Baskaran}\ and\ \citenamefont
  {Marchetti}(2008)}]{Baskaran2008c}%
  \BibitemOpen
  \bibfield  {author} {\bibinfo {author} {\bibfnamefont {A.}~\bibnamefont
  {Baskaran}}\ and\ \bibinfo {author} {\bibfnamefont {M.}~\bibnamefont
  {Marchetti}},\ }\href {\doibase 10.1103/PhysRevLett.101.268101} {\bibfield
  {journal} {\bibinfo  {journal} {Phys. Rev. Lett.}\ }\textbf {\bibinfo
  {volume} {101}},\ \bibinfo {pages} {1} (\bibinfo {year} {2008})}\BibitemShut
  {NoStop}%
\bibitem [{\citenamefont {Bertin}\ \emph {et~al.}(2006)\citenamefont {Bertin},
  \citenamefont {Droz},\ and\ \citenamefont {Gr\'{e}goire}}]{Bertin2006}%
  \BibitemOpen
  \bibfield  {author} {\bibinfo {author} {\bibfnamefont {E.}~\bibnamefont
  {Bertin}}, \bibinfo {author} {\bibfnamefont {M.}~\bibnamefont {Droz}}, \ and\
  \bibinfo {author} {\bibfnamefont {G.}~\bibnamefont {Gr\'{e}goire}},\ }\href
  {\doibase 10.1103/PhysRevE.74.022101} {\bibfield  {journal} {\bibinfo
  {journal} {Phys. Rev. E}\ }\textbf {\bibinfo {volume} {74}},\ \bibinfo
  {pages} {1} (\bibinfo {year} {2006})}\BibitemShut {NoStop}%
\bibitem [{\citenamefont {Bertin}\ \emph {et~al.}(2009)\citenamefont {Bertin},
  \citenamefont {Droz},\ and\ \citenamefont {Gr\'{e}goire}}]{Bertin2009}%
  \BibitemOpen
  \bibfield  {author} {\bibinfo {author} {\bibfnamefont {E.}~\bibnamefont
  {Bertin}}, \bibinfo {author} {\bibfnamefont {M.}~\bibnamefont {Droz}}, \ and\
  \bibinfo {author} {\bibfnamefont {G.}~\bibnamefont {Gr\'{e}goire}},\ }\href
  {\doibase 10.1088/1751-8113/42/44/445001} {\bibfield  {journal} {\bibinfo
  {journal} {J. Phys. A}\ }\textbf {\bibinfo {volume} {42}},\ \bibinfo {pages}
  {445001} (\bibinfo {year} {2009})}\BibitemShut {NoStop}%
\bibitem [{\citenamefont {Cisneros}\ \emph {et~al.}(2011)\citenamefont
  {Cisneros}, \citenamefont {Kessler}, \citenamefont {Ganguly},\ and\
  \citenamefont {Goldstein}}]{Cisneros2011a}%
  \BibitemOpen
  \bibfield  {author} {\bibinfo {author} {\bibfnamefont {L.}~\bibnamefont
  {Cisneros}}, \bibinfo {author} {\bibfnamefont {J.}~\bibnamefont {Kessler}},
  \bibinfo {author} {\bibfnamefont {S.}~\bibnamefont {Ganguly}}, \ and\
  \bibinfo {author} {\bibfnamefont {R.}~\bibnamefont {Goldstein}},\ }\href
  {\doibase 10.1103/PhysRevE.83.061907} {\bibfield  {journal} {\bibinfo
  {journal} {Phys. Rev. E}\ }\textbf {\bibinfo {volume} {83}},\ \bibinfo
  {pages} {061907} (\bibinfo {year} {2011})}\BibitemShut {NoStop}%
\bibitem [{\citenamefont {Ginelli}\ \emph {et~al.}(2010)\citenamefont
  {Ginelli}, \citenamefont {Peruani}, \citenamefont {B\"{a}r},\ and\
  \citenamefont {Chat\'{e}}}]{Ginelli2010a}%
  \BibitemOpen
  \bibfield  {author} {\bibinfo {author} {\bibfnamefont {F.}~\bibnamefont
  {Ginelli}}, \bibinfo {author} {\bibfnamefont {F.}~\bibnamefont {Peruani}},
  \bibinfo {author} {\bibfnamefont {M.}~\bibnamefont {B\"{a}r}}, \ and\
  \bibinfo {author} {\bibfnamefont {H.}~\bibnamefont {Chat\'{e}}},\ }\href
  {\doibase 10.1103/PhysRevLett.104.184502} {\bibfield  {journal} {\bibinfo
  {journal} {Phys. Rev. Lett.}\ }\textbf {\bibinfo {volume} {104}},\ \bibinfo
  {pages} {1} (\bibinfo {year} {2010})}\BibitemShut {NoStop}%
\bibitem [{\citenamefont {Head}\ \emph {et~al.}(2010)\citenamefont {Head},
  \citenamefont {Gompper},\ and\ \citenamefont {Briels}}]{Head2010a}%
  \BibitemOpen
  \bibfield  {author} {\bibinfo {author} {\bibfnamefont {D.~A.}\ \bibnamefont
  {Head}}, \bibinfo {author} {\bibfnamefont {G.}~\bibnamefont {Gompper}}, \
  and\ \bibinfo {author} {\bibfnamefont {W.~J.}\ \bibnamefont {Briels}},\
  }\href {\doibase 10.1039/b000000x} {\bibfield  {journal} {\bibinfo  {journal}
  {Soft Matter}\ ,\ \bibinfo {pages} {14}} (\bibinfo {year} {2010})},\ \Eprint
  {http://arxiv.org/abs/1009.1986} {arXiv:1009.1986} \BibitemShut {NoStop}%
\bibitem [{\citenamefont {Jia}\ \emph {et~al.}(2008)\citenamefont {Jia},
  \citenamefont {Karpeev}, \citenamefont {Aranson},\ and\ \citenamefont
  {Bates}}]{Jia2008a}%
  \BibitemOpen
  \bibfield  {author} {\bibinfo {author} {\bibfnamefont {Z.}~\bibnamefont
  {Jia}}, \bibinfo {author} {\bibfnamefont {D.}~\bibnamefont {Karpeev}},
  \bibinfo {author} {\bibfnamefont {I.~S.}\ \bibnamefont {Aranson}}, \ and\
  \bibinfo {author} {\bibfnamefont {P.~W.}\ \bibnamefont {Bates}},\ }\href
  {\doibase 10.1103/PhysRevE.77.051905} {\bibfield  {journal} {\bibinfo
  {journal} {Phys. Rev. E}\ }\textbf {\bibinfo {volume} {77}},\ \bibinfo
  {pages} {1} (\bibinfo {year} {2008})}\BibitemShut {NoStop}%
\bibitem [{\citenamefont {Peruani}\ \emph {et~al.}(2006)\citenamefont
  {Peruani}, \citenamefont {Deutsch},\ and\ \citenamefont
  {B\"{a}r}}]{Peruani2006a}%
  \BibitemOpen
  \bibfield  {author} {\bibinfo {author} {\bibfnamefont {F.}~\bibnamefont
  {Peruani}}, \bibinfo {author} {\bibfnamefont {A.}~\bibnamefont {Deutsch}}, \
  and\ \bibinfo {author} {\bibfnamefont {M.}~\bibnamefont {B\"{a}r}},\ }\href
  {\doibase 10.1103/PhysRevE.74.030904} {\bibfield  {journal} {\bibinfo
  {journal} {Phys. Rev. E}\ }\textbf {\bibinfo {volume} {74}},\ \bibinfo
  {pages} {1} (\bibinfo {year} {2006})}\BibitemShut {NoStop}%
\bibitem [{\citenamefont {Peruani}\ \emph
  {et~al.}(2011{\natexlab{a}})\citenamefont {Peruani}, \citenamefont {Klauss},
  \citenamefont {Deutsch},\ and\ \citenamefont {Voss-Boehme}}]{Peruani2011}%
  \BibitemOpen
  \bibfield  {author} {\bibinfo {author} {\bibfnamefont {F.}~\bibnamefont
  {Peruani}}, \bibinfo {author} {\bibfnamefont {T.}~\bibnamefont {Klauss}},
  \bibinfo {author} {\bibfnamefont {A.}~\bibnamefont {Deutsch}}, \ and\
  \bibinfo {author} {\bibfnamefont {A.}~\bibnamefont {Voss-Boehme}},\ }\href
  {\doibase 10.1103/PhysRevLett.106.128101} {\bibfield  {journal} {\bibinfo
  {journal} {Phys. Rev. Lett.}\ }\textbf {\bibinfo {volume} {106}},\ \bibinfo
  {pages} {1} (\bibinfo {year} {2011}{\natexlab{a}})}\BibitemShut {NoStop}%
\bibitem [{\citenamefont {Peruani}\ \emph
  {et~al.}(2011{\natexlab{b}})\citenamefont {Peruani}, \citenamefont
  {Schimansky-Geier},\ and\ \citenamefont {B\"{a}r}}]{Peruani2011a}%
  \BibitemOpen
  \bibfield  {author} {\bibinfo {author} {\bibfnamefont {F.}~\bibnamefont
  {Peruani}}, \bibinfo {author} {\bibfnamefont {L.}~\bibnamefont
  {Schimansky-Geier}}, \ and\ \bibinfo {author} {\bibfnamefont
  {M.}~\bibnamefont {B\"{a}r}},\ }\href {\doibase 10.1140/epjst/e2010-01349-1}
  {\bibfield  {journal} {\bibinfo  {journal} {Euro. Phys. Spec. Top.}\ }\textbf
  {\bibinfo {volume} {191}},\ \bibinfo {pages} {173} (\bibinfo {year}
  {2011}{\natexlab{b}})}\BibitemShut {NoStop}%
\bibitem [{\citenamefont {Saintillan}\ and\ \citenamefont
  {Shelley}(2011)}]{Saintillan2011}%
  \BibitemOpen
  \bibfield  {author} {\bibinfo {author} {\bibfnamefont {D.}~\bibnamefont
  {Saintillan}}\ and\ \bibinfo {author} {\bibfnamefont {M.~J.}\ \bibnamefont
  {Shelley}},\ }\href {\doibase 10.1098/rsif.2011.0355} {\bibfield  {journal}
  {\bibinfo  {journal} {J. R. Soc., Interface}\ ,\ \bibinfo {pages} {1}}
  (\bibinfo {year} {2011})}\BibitemShut {NoStop}%
\bibitem [{\citenamefont {Vicsek}\ \emph {et~al.}(1995)\citenamefont {Vicsek},
  \citenamefont {Czir\'{o}k}, \citenamefont {Ben-Jacob}, \citenamefont
  {Cohen},\ and\ \citenamefont {Shochet}}]{Vicsek1995a}%
  \BibitemOpen
  \bibfield  {author} {\bibinfo {author} {\bibfnamefont {T.}~\bibnamefont
  {Vicsek}}, \bibinfo {author} {\bibfnamefont {A.}~\bibnamefont {Czir\'{o}k}},
  \bibinfo {author} {\bibfnamefont {E.}~\bibnamefont {Ben-Jacob}}, \bibinfo
  {author} {\bibfnamefont {I.}~\bibnamefont {Cohen}}, \ and\ \bibinfo {author}
  {\bibfnamefont {O.}~\bibnamefont {Shochet}},\ }\href
  {http://link.aps.org/doi/10.1103/PhysRevLett.75.1226} {\bibfield  {journal}
  {\bibinfo  {journal} {Phys. Rev. Lett.}\ }\textbf {\bibinfo {volume} {75}},\
  \bibinfo {pages} {1226} (\bibinfo {year} {1995})}\BibitemShut {NoStop}%
\bibitem [{\citenamefont {Yang}\ \emph {et~al.}(2010)\citenamefont {Yang},
  \citenamefont {Marceau},\ and\ \citenamefont {Gompper}}]{Yang2010a}%
  \BibitemOpen
  \bibfield  {author} {\bibinfo {author} {\bibfnamefont {Y.}~\bibnamefont
  {Yang}}, \bibinfo {author} {\bibfnamefont {V.}~\bibnamefont {Marceau}}, \
  and\ \bibinfo {author} {\bibfnamefont {G.}~\bibnamefont {Gompper}},\ }\href
  {\doibase 10.1103/PhysRevE.82.031904} {\bibfield  {journal} {\bibinfo
  {journal} {Phys. Rev. E}\ }\textbf {\bibinfo {volume} {82}},\ \bibinfo
  {pages} {1} (\bibinfo {year} {2010})}\BibitemShut {NoStop}%
\bibitem [{\citenamefont {Chai}\ \emph {et~al.}(2011)\citenamefont {Chai},
  \citenamefont {Vlamakis},\ and\ \citenamefont {Kolter}}]{Chai2011}%
  \BibitemOpen
  \bibfield  {author} {\bibinfo {author} {\bibfnamefont {L.}~\bibnamefont
  {Chai}}, \bibinfo {author} {\bibfnamefont {H.}~\bibnamefont {Vlamakis}}, \
  and\ \bibinfo {author} {\bibfnamefont {R.}~\bibnamefont {Kolter}},\ }\href
  {\doibase 10.1557/mrs.2011.68} {\bibfield  {journal} {\bibinfo  {journal}
  {MRS Bull.}\ }\textbf {\bibinfo {volume} {36}},\ \bibinfo {pages} {374}
  (\bibinfo {year} {2011})}\BibitemShut {NoStop}%
\bibitem [{Sup()}]{SupplementalInformation}%
  \BibitemOpen
  \href@noop {} {}\bibinfo {note} {See supplemental material for this paper on
  arXiv Data Conservancy, or at the web page of A.B.
  (\url{http://people.brandeis.edu/~aparna/}) or M.H.
  (\url{http://www.brandeis.edu/departments/physics/hagan/index.html}).}\BibitemShut
  {Stop}%
\bibitem [{\citenamefont {Li}\ and\ \citenamefont {Tang}(2009)}]{Li2009}%
  \BibitemOpen
  \bibfield  {author} {\bibinfo {author} {\bibfnamefont {G.}~\bibnamefont
  {Li}}\ and\ \bibinfo {author} {\bibfnamefont {J.~X.}\ \bibnamefont {Tang}},\
  }\href {\doibase 10.1103/PhysRevLett.103.078101} {\bibfield  {journal}
  {\bibinfo  {journal} {Phys. Rev. Lett.}\ }\textbf {\bibinfo {volume} {103}},\
  \bibinfo {pages} {078101} (\bibinfo {year} {2009})}\BibitemShut {NoStop}%
\bibitem [{\citenamefont {Drescher}\ \emph {et~al.}(2011)\citenamefont
  {Drescher}, \citenamefont {Dunkel}, \citenamefont {Cisneros}, \citenamefont
  {Ganguly},\ and\ \citenamefont {Goldstein}}]{Drescher2011}%
  \BibitemOpen
  \bibfield  {author} {\bibinfo {author} {\bibfnamefont {K.}~\bibnamefont
  {Drescher}}, \bibinfo {author} {\bibfnamefont {J.}~\bibnamefont {Dunkel}},
  \bibinfo {author} {\bibfnamefont {L.~H.}\ \bibnamefont {Cisneros}}, \bibinfo
  {author} {\bibfnamefont {S.}~\bibnamefont {Ganguly}}, \ and\ \bibinfo
  {author} {\bibfnamefont {R.~E.}\ \bibnamefont {Goldstein}},\ }\href {\doibase
  10.1073/pnas.1019079108} {\bibfield  {journal} {\bibinfo  {journal} {Proc.
  Natl. Acad. Sci. U. S. A.}\ }\textbf {\bibinfo {volume} {108}},\ \bibinfo
  {pages} {10940} (\bibinfo {year} {2011})}\BibitemShut {NoStop}%
\bibitem [{\citenamefont {Tailleur}\ and\ \citenamefont
  {Cates}(2008)}]{Tailleur2008}%
  \BibitemOpen
  \bibfield  {author} {\bibinfo {author} {\bibfnamefont {J.}~\bibnamefont
  {Tailleur}}\ and\ \bibinfo {author} {\bibfnamefont {M.~E.}\ \bibnamefont
  {Cates}},\ }\href@noop {} {\bibfield  {journal} {\bibinfo  {journal} {Phys.
  Rev. Lett.}\ }\textbf {\bibinfo {volume} {100}},\ \bibinfo {pages} {218103}
  (\bibinfo {year} {2008})}\BibitemShut {NoStop}%
\bibitem [{\citenamefont {Cates}\ \emph {et~al.}(2010)\citenamefont {Cates},
  \citenamefont {Marenduzzo}, \citenamefont {Pagonabarraga},\ and\
  \citenamefont {Tailleur}}]{Cates2010}%
  \BibitemOpen
  \bibfield  {author} {\bibinfo {author} {\bibfnamefont {M.~E.}\ \bibnamefont
  {Cates}}, \bibinfo {author} {\bibfnamefont {D.}~\bibnamefont {Marenduzzo}},
  \bibinfo {author} {\bibfnamefont {I.}~\bibnamefont {Pagonabarraga}}, \ and\
  \bibinfo {author} {\bibfnamefont {J.}~\bibnamefont {Tailleur}},\ }\href@noop
  {} {\bibfield  {journal} {\bibinfo  {journal} {Proc. Natl. Acad. Sci. U. S.
  A.}\ }\textbf {\bibinfo {volume} {107}},\ \bibinfo {pages} {11715} (\bibinfo
  {year} {2010})}\BibitemShut {NoStop}%
\bibitem [{\citenamefont {Thompson}\ \emph {et~al.}(2011)\citenamefont
  {Thompson}, \citenamefont {Tailleur}, \citenamefont {Cates},\ and\
  \citenamefont {Blythe}}]{Thompson2011}%
  \BibitemOpen
  \bibfield  {author} {\bibinfo {author} {\bibfnamefont {A.~G.}\ \bibnamefont
  {Thompson}}, \bibinfo {author} {\bibfnamefont {J.}~\bibnamefont {Tailleur}},
  \bibinfo {author} {\bibfnamefont {M.~E.}\ \bibnamefont {Cates}}, \ and\
  \bibinfo {author} {\bibfnamefont {R.~A.}\ \bibnamefont {Blythe}},\ }\href
  {http://stacks.iop.org/1742-5468/2011/i=02/a=P02029} {\bibfield  {journal}
  {\bibinfo  {journal} {J. Stat. Mech. Theor. Exp.}\ }\textbf {\bibinfo
  {volume} {2011}},\ \bibinfo {pages} {P02029} (\bibinfo {year}
  {2011})}\BibitemShut {NoStop}%
\bibitem [{\citenamefont {Weeks}\ \emph {et~al.}(1971)\citenamefont {Weeks},
  \citenamefont {D.},\ and\ \citenamefont {Andersen}}]{Weeks1971a}%
  \BibitemOpen
  \bibfield  {author} {\bibinfo {author} {\bibfnamefont {J.~D.}\ \bibnamefont
  {Weeks}}, \bibinfo {author} {\bibfnamefont {C.}~\bibnamefont {D.}}, \ and\
  \bibinfo {author} {\bibfnamefont {H.~C.}\ \bibnamefont {Andersen}},\ }\href
  {\doibase 10.1063/1.1674820} {\bibfield  {journal} {\bibinfo  {journal} {J.
  Chem. Phys.}\ }\textbf {\bibinfo {volume} {54}},\ \bibinfo {pages} {5237}
  (\bibinfo {year} {1971})}\BibitemShut {NoStop}%
\bibitem [{\citenamefont {Tao}\ \emph {et~al.}(2005)\citenamefont {Tao},
  \citenamefont {den Otter}, \citenamefont {Padding}, \citenamefont {Dhont},\
  and\ \citenamefont {Briels}}]{Tao2005a}%
  \BibitemOpen
  \bibfield  {author} {\bibinfo {author} {\bibfnamefont {Y.-G.}\ \bibnamefont
  {Tao}}, \bibinfo {author} {\bibfnamefont {W.~K.}\ \bibnamefont {den Otter}},
  \bibinfo {author} {\bibfnamefont {J.~T.}\ \bibnamefont {Padding}}, \bibinfo
  {author} {\bibfnamefont {J.~K.~G.}\ \bibnamefont {Dhont}}, \ and\ \bibinfo
  {author} {\bibfnamefont {W.~J.}\ \bibnamefont {Briels}},\ }\href {\doibase
  10.1063/1.1940031} {\bibfield  {journal} {\bibinfo  {journal} {J. Chem.
  Phys.}\ }\textbf {\bibinfo {volume} {122}},\ \bibinfo {pages} {244903}
  (\bibinfo {year} {2005})}\BibitemShut {NoStop}%
\bibitem [{\citenamefont {Branka}\ and\ \citenamefont
  {Heyes}(1999)}]{Branka1999}%
  \BibitemOpen
  \bibfield  {author} {\bibinfo {author} {\bibfnamefont {A.}~\bibnamefont
  {Branka}}\ and\ \bibinfo {author} {\bibfnamefont {D.}~\bibnamefont {Heyes}},\
  }\href@noop {} {\bibfield  {journal} {\bibinfo  {journal} {Phys. Rev. E}\
  }\textbf {\bibinfo {volume} {60}},\ \bibinfo {pages} {2381} (\bibinfo {year}
  {1999})}\BibitemShut {NoStop}%
\bibitem [{\citenamefont {Heyes}\ and\ \citenamefont
  {Branka}(2000)}]{Heyes2000}%
  \BibitemOpen
  \bibfield  {author} {\bibinfo {author} {\bibfnamefont {D.}~\bibnamefont
  {Heyes}}\ and\ \bibinfo {author} {\bibfnamefont {A.}~\bibnamefont {Branka}},\
  }\href@noop {} {\bibfield  {journal} {\bibinfo  {journal} {Mol. Phys.}\
  }\textbf {\bibinfo {volume} {98}},\ \bibinfo {pages} {1949} (\bibinfo {year}
  {2000})}\BibitemShut {NoStop}%
\bibitem [{Note1()}]{Note1}%
  \BibitemOpen
  \bibinfo {note} {The actual timestep used for a given iteration is $\delta
  t/n$, where $n$ is initially set to one. We mark an integration error if the
  force on a rod is such that it will be moved more than $0.6/n$ rod diameters,
  rotated more than $4\sigma /\ell n$ radians, or if the lines representing two
  rods overlap. Such events are inevitable in Brownian dynamics at high density
  at any timestep for a sufficiently long run because of the Gaussian noise. If
  an error is recorded, we revert back to a checkpoint saved every 10,000
  timesteps and increment $n$. We decrement $n$ after 10,000 successful
  timesteps.}\BibitemShut {Stop}%
\bibitem [{\citenamefont {Vissers}\ \emph {et~al.}(2011)\citenamefont
  {Vissers}, \citenamefont {Wysocki}, \citenamefont {Rex}, \citenamefont
  {L\"{o}wen}, \citenamefont {Royall}, \citenamefont {Imhof},\ and\
  \citenamefont {van Blaaderen}}]{Vissers2011}%
  \BibitemOpen
  \bibfield  {author} {\bibinfo {author} {\bibfnamefont {T.}~\bibnamefont
  {Vissers}}, \bibinfo {author} {\bibfnamefont {A.}~\bibnamefont {Wysocki}},
  \bibinfo {author} {\bibfnamefont {M.}~\bibnamefont {Rex}}, \bibinfo {author}
  {\bibfnamefont {H.}~\bibnamefont {L\"{o}wen}}, \bibinfo {author}
  {\bibfnamefont {C.~P.}\ \bibnamefont {Royall}}, \bibinfo {author}
  {\bibfnamefont {A.}~\bibnamefont {Imhof}}, \ and\ \bibinfo {author}
  {\bibfnamefont {A.}~\bibnamefont {van Blaaderen}},\ }\href {\doibase
  10.1039/c0sm01343a} {\bibfield  {journal} {\bibinfo  {journal} {Soft Matter}\
  }\textbf {\bibinfo {volume} {7}},\ \bibinfo {pages} {2352} (\bibinfo {year}
  {2011})}\BibitemShut {NoStop}%
\bibitem [{\citenamefont {Valiveti}\ and\ \citenamefont
  {Koch}(1999)}]{Valiveti1999}%
  \BibitemOpen
  \bibfield  {author} {\bibinfo {author} {\bibfnamefont {P.}~\bibnamefont
  {Valiveti}}\ and\ \bibinfo {author} {\bibfnamefont {D.~L.}\ \bibnamefont
  {Koch}},\ }\href {\doibase 10.1063/1.870188} {\bibfield  {journal} {\bibinfo
  {journal} {Phys. Fluids}\ }\textbf {\bibinfo {volume} {11}},\ \bibinfo
  {pages} {3283} (\bibinfo {year} {1999})}\BibitemShut {NoStop}%
\bibitem [{\citenamefont {Batchelor}\ and\ \citenamefont
  {Vanrensburg}(1986)}]{Batchelor1986}%
  \BibitemOpen
  \bibfield  {author} {\bibinfo {author} {\bibfnamefont {G.~K.}\ \bibnamefont
  {Batchelor}}\ and\ \bibinfo {author} {\bibfnamefont {R.~W.~J.}\ \bibnamefont
  {Vanrensburg}},\ }\href@noop {} {\bibfield  {journal} {\bibinfo  {journal}
  {J. Fluid Mech.}\ }\textbf {\bibinfo {volume} {166}},\ \bibinfo {pages} {379}
  (\bibinfo {year} {1986})}\BibitemShut {NoStop}%
\bibitem [{\citenamefont {Cox}(1990)}]{Cox1990}%
  \BibitemOpen
  \bibfield  {author} {\bibinfo {author} {\bibfnamefont {R.~G.}\ \bibnamefont
  {Cox}},\ }\href@noop {} {\bibfield  {journal} {\bibinfo  {journal} {Int. J.
  Multiphase Flow}\ }\textbf {\bibinfo {volume} {16}},\ \bibinfo {pages} {617}
  (\bibinfo {year} {1990})}\BibitemShut {NoStop}%
\bibitem [{\citenamefont {Couzin}\ and\ \citenamefont
  {Franks}(2003)}]{Couzin2003}%
  \BibitemOpen
  \bibfield  {author} {\bibinfo {author} {\bibfnamefont {I.~D.}\ \bibnamefont
  {Couzin}}\ and\ \bibinfo {author} {\bibfnamefont {N.~R.}\ \bibnamefont
  {Franks}},\ }\href@noop {} {\bibfield  {journal} {\bibinfo  {journal} {Proc.
  R. Soc. B}\ }\textbf {\bibinfo {volume} {270}},\ \bibinfo {pages} {139}
  (\bibinfo {year} {2003})}\BibitemShut {NoStop}%
\bibitem [{\citenamefont {Milgram}\ and\ \citenamefont
  {Toch}(1969)}]{Milgram1969}%
  \BibitemOpen
  \bibfield  {author} {\bibinfo {author} {\bibfnamefont {S.}~\bibnamefont
  {Milgram}}\ and\ \bibinfo {author} {\bibfnamefont {H.}~\bibnamefont {Toch}},\
  }\enquote {\bibinfo {title} {The handbook of social psychology},}\ \
  (\bibinfo  {publisher} {Addison-Wesley},\ \bibinfo {year} {1969})\ Chap.\
  \bibinfo {chapter} {Collective behaviour: crowds and social
  movements}\BibitemShut {NoStop}%
\bibitem [{\citenamefont {Delhommelle}(2005)}]{Delhommelle2005}%
  \BibitemOpen
  \bibfield  {author} {\bibinfo {author} {\bibfnamefont {J.}~\bibnamefont
  {Delhommelle}},\ }\href {\doibase 10.1103/PhysRevE.71.016705} {\bibfield
  {journal} {\bibinfo  {journal} {Phys. Rev. E}\ }\textbf {\bibinfo {volume}
  {71}},\ \bibinfo {pages} {016705} (\bibinfo {year} {2005})}\BibitemShut
  {NoStop}%
\bibitem [{\citenamefont {Dzubiella}\ \emph {et~al.}(2002)\citenamefont
  {Dzubiella}, \citenamefont {Hoffmann},\ and\ \citenamefont
  {L\"owen}}]{Dzubiella2002}%
  \BibitemOpen
  \bibfield  {author} {\bibinfo {author} {\bibfnamefont {J.}~\bibnamefont
  {Dzubiella}}, \bibinfo {author} {\bibfnamefont {G.~P.}\ \bibnamefont
  {Hoffmann}}, \ and\ \bibinfo {author} {\bibfnamefont {H.}~\bibnamefont
  {L\"owen}},\ }\href {\doibase 10.1103/PhysRevE.65.021402} {\bibfield
  {journal} {\bibinfo  {journal} {Phys. Rev. E}\ }\textbf {\bibinfo {volume}
  {65}},\ \bibinfo {pages} {021402} (\bibinfo {year} {2002})}\BibitemShut
  {NoStop}%
\bibitem [{\citenamefont {Loewen}(2010)}]{Loewen2010}%
  \BibitemOpen
  \bibfield  {author} {\bibinfo {author} {\bibfnamefont {H.}~\bibnamefont
  {Loewen}},\ }\href@noop {} {\bibfield  {journal} {\bibinfo  {journal} {Soft
  Matter}\ }\textbf {\bibinfo {volume} {6}},\ \bibinfo {pages} {3133} (\bibinfo
  {year} {2010})}\BibitemShut {NoStop}%
\bibitem [{\citenamefont {Rex}\ and\ \citenamefont {Lowen}(2007)}]{Rex2007}%
  \BibitemOpen
  \bibfield  {author} {\bibinfo {author} {\bibfnamefont {M.}~\bibnamefont
  {Rex}}\ and\ \bibinfo {author} {\bibfnamefont {H.}~\bibnamefont {Lowen}},\
  }\href@noop {} {\bibfield  {journal} {\bibinfo  {journal} {Phys. Rev. E}\
  }\textbf {\bibinfo {volume} {75}},\ \bibinfo {pages} {051402} (\bibinfo
  {year} {2007})}\BibitemShut {NoStop}%
\bibitem [{\citenamefont {S\"utterlin}\ \emph {et~al.}(2009)\citenamefont
  {S\"utterlin}, \citenamefont {Wysocki}, \citenamefont {Ivlev}, \citenamefont
  {R\"ath}, \citenamefont {Thomas}, \citenamefont {Rubin-Zuzic}, \citenamefont
  {Goedheer}, \citenamefont {Fortov}, \citenamefont {Lipaev}, \citenamefont
  {Molotkov}, \citenamefont {Petrov}, \citenamefont {Morfill},\ and\
  \citenamefont {L\"owen}}]{Sutterlin2009}%
  \BibitemOpen
  \bibfield  {author} {\bibinfo {author} {\bibfnamefont {K.~R.}\ \bibnamefont
  {S\"utterlin}}, \bibinfo {author} {\bibfnamefont {A.}~\bibnamefont
  {Wysocki}}, \bibinfo {author} {\bibfnamefont {A.~V.}\ \bibnamefont {Ivlev}},
  \bibinfo {author} {\bibfnamefont {C.}~\bibnamefont {R\"ath}}, \bibinfo
  {author} {\bibfnamefont {H.~M.}\ \bibnamefont {Thomas}}, \bibinfo {author}
  {\bibfnamefont {M.}~\bibnamefont {Rubin-Zuzic}}, \bibinfo {author}
  {\bibfnamefont {W.~J.}\ \bibnamefont {Goedheer}}, \bibinfo {author}
  {\bibfnamefont {V.~E.}\ \bibnamefont {Fortov}}, \bibinfo {author}
  {\bibfnamefont {A.~M.}\ \bibnamefont {Lipaev}}, \bibinfo {author}
  {\bibfnamefont {V.~I.}\ \bibnamefont {Molotkov}}, \bibinfo {author}
  {\bibfnamefont {O.~F.}\ \bibnamefont {Petrov}}, \bibinfo {author}
  {\bibfnamefont {G.~E.}\ \bibnamefont {Morfill}}, \ and\ \bibinfo {author}
  {\bibfnamefont {H.}~\bibnamefont {L\"owen}},\ }\href {\doibase
  10.1103/PhysRevLett.102.085003} {\bibfield  {journal} {\bibinfo  {journal}
  {Phys. Rev. Lett.}\ }\textbf {\bibinfo {volume} {102}},\ \bibinfo {pages}
  {085003} (\bibinfo {year} {2009})}\BibitemShut {NoStop}%
\bibitem [{\citenamefont {Wysocki}\ and\ \citenamefont
  {Loewen}(2011)}]{Wysocki2011}%
  \BibitemOpen
  \bibfield  {author} {\bibinfo {author} {\bibfnamefont {A.}~\bibnamefont
  {Wysocki}}\ and\ \bibinfo {author} {\bibfnamefont {H.}~\bibnamefont
  {Loewen}},\ }\href@noop {} {\bibfield  {journal} {\bibinfo  {journal} {J.
  Phys. Cond. Matter}\ }\textbf {\bibinfo {volume} {23}},\ \bibinfo {pages}
  {284117} (\bibinfo {year} {2011})}\BibitemShut {NoStop}%
\bibitem [{Note2()}]{Note2}%
  \BibitemOpen
  \bibinfo {note} {The bands did not break down within our finite simulation
  time for any of our systems with Pe = 10, but we anticipate that breakdown
  would eventually occur for sufficiently long simulations.}\BibitemShut
  {Stop}%
\bibitem [{\citenamefont {Narayan}\ \emph {et~al.}(2006)\citenamefont
  {Narayan}, \citenamefont {Menon},\ and\ \citenamefont
  {Ramaswamy}}]{Narayan2006}%
  \BibitemOpen
  \bibfield  {author} {\bibinfo {author} {\bibfnamefont {V.}~\bibnamefont
  {Narayan}}, \bibinfo {author} {\bibfnamefont {N.}~\bibnamefont {Menon}}, \
  and\ \bibinfo {author} {\bibfnamefont {S.}~\bibnamefont {Ramaswamy}},\ }\href
  {\doibase 10.1088/1742-5468/2006/01/P01005} {\bibfield  {journal} {\bibinfo
  {journal} {J. Stat. Mech.}\ }\textbf {\bibinfo {volume} {2006}},\ \bibinfo
  {pages} {1} (\bibinfo {year} {2006})}\BibitemShut {NoStop}%
\bibitem [{\citenamefont {Massey}\ and\ \citenamefont
  {Denton}(1988)}]{Massey1988}%
  \BibitemOpen
  \bibfield  {author} {\bibinfo {author} {\bibfnamefont {D.~S.}\ \bibnamefont
  {Massey}}\ and\ \bibinfo {author} {\bibfnamefont {N.~a.}\ \bibnamefont
  {Denton}},\ }\href {\doibase 10.2307/2579183} {\bibfield  {journal} {\bibinfo
   {journal} {Social Forces}\ }\textbf {\bibinfo {volume} {67}},\ \bibinfo
  {pages} {281} (\bibinfo {year} {1988})}\BibitemShut {NoStop}%
\bibitem [{\citenamefont {Chat\'{e}}\ \emph
  {et~al.}(2008{\natexlab{b}})\citenamefont {Chat\'{e}}, \citenamefont
  {Ginelli},\ and\ \citenamefont {Raynaud}}]{Chate2008b}%
  \BibitemOpen
  \bibfield  {author} {\bibinfo {author} {\bibfnamefont {H.}~\bibnamefont
  {Chat\'{e}}}, \bibinfo {author} {\bibfnamefont {F.}~\bibnamefont {Ginelli}},
  \ and\ \bibinfo {author} {\bibfnamefont {F.}~\bibnamefont {Raynaud}},\ }\href
  {\doibase 10.1103/PhysRevE.77.046113} {\bibfield  {journal} {\bibinfo
  {journal} {Phys. Rev. E}\ }\textbf {\bibinfo {volume} {77}},\ \bibinfo
  {pages} {1} (\bibinfo {year} {2008}{\natexlab{b}})}\BibitemShut {NoStop}%
\bibitem [{\citenamefont {ten Hagen}\ \emph {et~al.}(2011)\citenamefont {ten
  Hagen}, \citenamefont {van Teeffelen},\ and\ \citenamefont
  {Lowen}}]{Hagen2011}%
  \BibitemOpen
  \bibfield  {author} {\bibinfo {author} {\bibfnamefont {B.}~\bibnamefont {ten
  Hagen}}, \bibinfo {author} {\bibfnamefont {S.}~\bibnamefont {van Teeffelen}},
  \ and\ \bibinfo {author} {\bibfnamefont {H.}~\bibnamefont {Lowen}},\
  }\href@noop {} {\bibfield  {journal} {\bibinfo  {journal} {J. Phys.: Condens.
  Matter}\ }\textbf {\bibinfo {volume} {23}},\ \bibinfo {pages} {194119}
  (\bibinfo {year} {2011})}\BibitemShut {NoStop}%
\bibitem [{\citenamefont {Prevost}\ \emph {et~al.}(2002)\citenamefont
  {Prevost}, \citenamefont {Egolf},\ and\ \citenamefont
  {Urbach}}]{Prevost2002}%
  \BibitemOpen
  \bibfield  {author} {\bibinfo {author} {\bibfnamefont {A.}~\bibnamefont
  {Prevost}}, \bibinfo {author} {\bibfnamefont {D.}~\bibnamefont {Egolf}}, \
  and\ \bibinfo {author} {\bibfnamefont {J.}~\bibnamefont {Urbach}},\ }\href
  {\doibase 10.1103/PhysRevLett.89.084301} {\bibfield  {journal} {\bibinfo
  {journal} {Phys. Rev. Lett.}\ }\textbf {\bibinfo {volume} {89}},\ \bibinfo
  {pages} {1} (\bibinfo {year} {2002})}\BibitemShut {NoStop}%
\bibitem [{\citenamefont {Blair}\ \emph {et~al.}(2003)\citenamefont {Blair},
  \citenamefont {Neicu},\ and\ \citenamefont {Kudrolli}}]{Blair2003}%
  \BibitemOpen
  \bibfield  {author} {\bibinfo {author} {\bibfnamefont {D.}~\bibnamefont
  {Blair}}, \bibinfo {author} {\bibfnamefont {T.}~\bibnamefont {Neicu}}, \ and\
  \bibinfo {author} {\bibfnamefont {a.}~\bibnamefont {Kudrolli}},\ }\href
  {\doibase 10.1103/PhysRevE.67.031303} {\bibfield  {journal} {\bibinfo
  {journal} {Phys. Rev. E}\ }\textbf {\bibinfo {volume} {67}},\ \bibinfo
  {pages} {1} (\bibinfo {year} {2003})}\BibitemShut {NoStop}%
\bibitem [{\citenamefont {Aranson}\ \emph {et~al.}(2007)\citenamefont
  {Aranson}, \citenamefont {Volfson},\ and\ \citenamefont
  {Tsimring}}]{Aranson2007}%
  \BibitemOpen
  \bibfield  {author} {\bibinfo {author} {\bibfnamefont {I.}~\bibnamefont
  {Aranson}}, \bibinfo {author} {\bibfnamefont {D.}~\bibnamefont {Volfson}}, \
  and\ \bibinfo {author} {\bibfnamefont {L.}~\bibnamefont {Tsimring}},\ }\href
  {\doibase 10.1103/PhysRevE.75.051301} {\bibfield  {journal} {\bibinfo
  {journal} {Phys. Rev. E}\ }\textbf {\bibinfo {volume} {75}},\ \bibinfo
  {pages} {1} (\bibinfo {year} {2007})}\BibitemShut {NoStop}%
\bibitem [{\citenamefont {Galanis}\ \emph {et~al.}(2006)\citenamefont
  {Galanis}, \citenamefont {Harries}, \citenamefont {Sackett}, \citenamefont
  {Losert},\ and\ \citenamefont {Nossal}}]{Galanis2006}%
  \BibitemOpen
  \bibfield  {author} {\bibinfo {author} {\bibfnamefont {J.}~\bibnamefont
  {Galanis}}, \bibinfo {author} {\bibfnamefont {D.}~\bibnamefont {Harries}},
  \bibinfo {author} {\bibfnamefont {D.}~\bibnamefont {Sackett}}, \bibinfo
  {author} {\bibfnamefont {W.}~\bibnamefont {Losert}}, \ and\ \bibinfo {author}
  {\bibfnamefont {R.}~\bibnamefont {Nossal}},\ }\href {\doibase
  10.1103/PhysRevLett.96.028002} {\bibfield  {journal} {\bibinfo  {journal}
  {Phys. Rev. Lett.}\ }\textbf {\bibinfo {volume} {96}},\ \bibinfo {pages} {5}
  (\bibinfo {year} {2006})}\BibitemShut {NoStop}%
\bibitem [{\citenamefont {Daniels}\ and\ \citenamefont
  {Durian}(2011)}]{Daniels2011}%
  \BibitemOpen
  \bibfield  {author} {\bibinfo {author} {\bibfnamefont {L.}~\bibnamefont
  {Daniels}}\ and\ \bibinfo {author} {\bibfnamefont {D.}~\bibnamefont
  {Durian}},\ }\href {\doibase 10.1103/PhysRevE.83.061304} {\bibfield
  {journal} {\bibinfo  {journal} {Phys. Rev. E}\ }\textbf {\bibinfo {volume}
  {83}},\ \bibinfo {pages} {1} (\bibinfo {year} {2011})}\BibitemShut {NoStop}%
\bibitem [{\citenamefont {Paxton}\ \emph {et~al.}(2004)\citenamefont {Paxton},
  \citenamefont {Kistler}, \citenamefont {Olmeda}, \citenamefont {Sen},
  \citenamefont {St~Angelo}, \citenamefont {Cao}, \citenamefont {Mallouk},
  \citenamefont {Lammert},\ and\ \citenamefont {Crespi}}]{Paxton2004}%
  \BibitemOpen
  \bibfield  {author} {\bibinfo {author} {\bibfnamefont {W.~F.}\ \bibnamefont
  {Paxton}}, \bibinfo {author} {\bibfnamefont {K.~C.}\ \bibnamefont {Kistler}},
  \bibinfo {author} {\bibfnamefont {C.~C.}\ \bibnamefont {Olmeda}}, \bibinfo
  {author} {\bibfnamefont {A.}~\bibnamefont {Sen}}, \bibinfo {author}
  {\bibfnamefont {S.~K.}\ \bibnamefont {St~Angelo}}, \bibinfo {author}
  {\bibfnamefont {Y.~Y.}\ \bibnamefont {Cao}}, \bibinfo {author} {\bibfnamefont
  {T.~E.}\ \bibnamefont {Mallouk}}, \bibinfo {author} {\bibfnamefont {P.~E.}\
  \bibnamefont {Lammert}}, \ and\ \bibinfo {author} {\bibfnamefont {V.~H.}\
  \bibnamefont {Crespi}},\ }\href@noop {} {\bibfield  {journal} {\bibinfo
  {journal} {J. Am. Chem. Soc.}\ }\textbf {\bibinfo {volume} {126}},\ \bibinfo
  {pages} {13424} (\bibinfo {year} {2004})}\BibitemShut {NoStop}%
\bibitem [{\citenamefont {Palacci}\ \emph {et~al.}(2010)\citenamefont
  {Palacci}, \citenamefont {Cottin-Bizonne}, \citenamefont {Ybert},\ and\
  \citenamefont {Bocquet}}]{Palacci2010}%
  \BibitemOpen
  \bibfield  {author} {\bibinfo {author} {\bibfnamefont {J.}~\bibnamefont
  {Palacci}}, \bibinfo {author} {\bibfnamefont {C.}~\bibnamefont
  {Cottin-Bizonne}}, \bibinfo {author} {\bibfnamefont {C.}~\bibnamefont
  {Ybert}}, \ and\ \bibinfo {author} {\bibfnamefont {L.}~\bibnamefont
  {Bocquet}},\ }\href@noop {} {\bibfield  {journal} {\bibinfo  {journal} {Phys.
  Rev. Lett.}\ }\textbf {\bibinfo {volume} {105}},\ \bibinfo {pages} {088304}
  (\bibinfo {year} {2010})}\BibitemShut {NoStop}%
\bibitem [{\citenamefont {Hong}\ \emph {et~al.}(2007)\citenamefont {Hong},
  \citenamefont {Blackman}, \citenamefont {Kopp}, \citenamefont {Sen},\ and\
  \citenamefont {Velegol}}]{Hong2007}%
  \BibitemOpen
  \bibfield  {author} {\bibinfo {author} {\bibfnamefont {Y.}~\bibnamefont
  {Hong}}, \bibinfo {author} {\bibfnamefont {N.~M.~K.}\ \bibnamefont
  {Blackman}}, \bibinfo {author} {\bibfnamefont {N.~D.}\ \bibnamefont {Kopp}},
  \bibinfo {author} {\bibfnamefont {A.}~\bibnamefont {Sen}}, \ and\ \bibinfo
  {author} {\bibfnamefont {D.}~\bibnamefont {Velegol}},\ }\href@noop {}
  {\bibfield  {journal} {\bibinfo  {journal} {Phys. Rev. Lett.}\ }\textbf
  {\bibinfo {volume} {99}},\ \bibinfo {pages} {178103} (\bibinfo {year}
  {2007})}\BibitemShut {NoStop}%
\bibitem [{\citenamefont {Ozin}\ \emph {et~al.}(2010)\citenamefont {Ozin},
  \citenamefont {Valadares}, \citenamefont {Tao}, \citenamefont {Zacharia},
  \citenamefont {Kitaev}, \citenamefont {Galembeck},\ and\ \citenamefont
  {Kapral}}]{Ozin2010}%
  \BibitemOpen
  \bibfield  {author} {\bibinfo {author} {\bibfnamefont {G.~A.}\ \bibnamefont
  {Ozin}}, \bibinfo {author} {\bibfnamefont {L.~F.}\ \bibnamefont {Valadares}},
  \bibinfo {author} {\bibfnamefont {Y.~G.}\ \bibnamefont {Tao}}, \bibinfo
  {author} {\bibfnamefont {N.~S.}\ \bibnamefont {Zacharia}}, \bibinfo {author}
  {\bibfnamefont {V.}~\bibnamefont {Kitaev}}, \bibinfo {author} {\bibfnamefont
  {F.}~\bibnamefont {Galembeck}}, \ and\ \bibinfo {author} {\bibfnamefont
  {R.}~\bibnamefont {Kapral}},\ }\href@noop {} {\bibfield  {journal} {\bibinfo
  {journal} {Small}\ }\textbf {\bibinfo {volume} {6}},\ \bibinfo {pages} {565}
  (\bibinfo {year} {2010})}\BibitemShut {NoStop}%
\bibitem [{\citenamefont {Thakur}\ and\ \citenamefont
  {Kapral}(2011)}]{Thakur2011}%
  \BibitemOpen
  \bibfield  {author} {\bibinfo {author} {\bibfnamefont {S.}~\bibnamefont
  {Thakur}}\ and\ \bibinfo {author} {\bibfnamefont {R.}~\bibnamefont
  {Kapral}},\ }\href@noop {} {\bibfield  {journal} {\bibinfo  {journal} {J.
  Chem. Phys.}\ }\textbf {\bibinfo {volume} {135}},\ \bibinfo {pages} {024509}
  (\bibinfo {year} {2011})}\BibitemShut {NoStop}%
\bibitem [{\citenamefont {Mi\~no}\ \emph {et~al.}(2011)\citenamefont {Mi\~no},
  \citenamefont {Mallouk}, \citenamefont {Darnige}, \citenamefont {Hoyos},
  \citenamefont {Dauchet}, \citenamefont {Dunstan}, \citenamefont {Soto},
  \citenamefont {Wang}, \citenamefont {Rousselet},\ and\ \citenamefont
  {Clement}}]{Mino2011}%
  \BibitemOpen
  \bibfield  {author} {\bibinfo {author} {\bibfnamefont {G.}~\bibnamefont
  {Mi\~no}}, \bibinfo {author} {\bibfnamefont {T.~E.}\ \bibnamefont {Mallouk}},
  \bibinfo {author} {\bibfnamefont {T.}~\bibnamefont {Darnige}}, \bibinfo
  {author} {\bibfnamefont {M.}~\bibnamefont {Hoyos}}, \bibinfo {author}
  {\bibfnamefont {J.}~\bibnamefont {Dauchet}}, \bibinfo {author} {\bibfnamefont
  {J.}~\bibnamefont {Dunstan}}, \bibinfo {author} {\bibfnamefont
  {R.}~\bibnamefont {Soto}}, \bibinfo {author} {\bibfnamefont {Y.}~\bibnamefont
  {Wang}}, \bibinfo {author} {\bibfnamefont {A.}~\bibnamefont {Rousselet}}, \
  and\ \bibinfo {author} {\bibfnamefont {E.}~\bibnamefont {Clement}},\ }\href
  {\doibase 10.1103/PhysRevLett.106.048102} {\bibfield  {journal} {\bibinfo
  {journal} {Phys. Rev. Lett.}\ }\textbf {\bibinfo {volume} {106}},\ \bibinfo
  {pages} {048102} (\bibinfo {year} {2011})}\BibitemShut {NoStop}%
\bibitem [{\citenamefont {Cogan}\ and\ \citenamefont
  {Wolgemuth}(2011)}]{Cogan2011}%
  \BibitemOpen
  \bibfield  {author} {\bibinfo {author} {\bibfnamefont {N.~G.}\ \bibnamefont
  {Cogan}}\ and\ \bibinfo {author} {\bibfnamefont {C.~W.}\ \bibnamefont
  {Wolgemuth}},\ }\href {\doibase 10.1007/s11538-010-9536-1} {\bibfield
  {journal} {\bibinfo  {journal} {Bull. Math. Bio.}\ }\textbf {\bibinfo
  {volume} {73}},\ \bibinfo {pages} {212} (\bibinfo {year} {2011})}\BibitemShut
  {NoStop}%
\end{thebibliography}%

\end{document}